\documentclass[twocolumn,aps,pra,showpacs,superscriptaddress,longbibliography,10pt]{revtex4-2}
\usepackage[colorlinks=true, allcolors=blue, citecolor=blue]{hyperref}
\usepackage{mathrsfs}
\usepackage{epsfig}
\usepackage{graphicx}
\usepackage{amsfonts}
\usepackage{amssymb}
\usepackage{amsmath}
\usepackage{dcolumn}
\usepackage{bm}
\usepackage{xcolor}
\usepackage{braket}
\usepackage{units}
\usepackage{xspace}
\usepackage{bbold}
\usepackage{orcidlink}
\usepackage{tabularx}
\usepackage{adjustbox}
\usepackage{booktabs}
\usepackage{bbding}
\usepackage{comment}
\usepackage[]{hyperref}

\newcommand*\midpoint[1]{\overline{#1}}
\newcommand{\ANU}{School of Physics, Anhui University, Hefei 230601, China}
\newcommand{\NU}{School of Physics and Technology, Nantong University, Nantong, 226019, China}
\newcommand{\JSU}{Department of Physics, Jiangsu University, Zhenjiang, Jiangsu 212013, China}
\newcommand{\ZJNU}{Department of Physics, Zhejiang Normal University, Jinhua 321004, China}

\begin{document}
\title{Decoherence across phase-space scales: From compass states to general quantum states}

\author{Naeem Akhtar}
\affiliation{\ANU}
\author{Jia-Xin Peng}
\email{JiaXinPeng@ntu.edu.cn}
\affiliation{\NU}
\author{Tan Hailin}
\email{tanhailin6626@outlook.com}
\affiliation{\ZJNU}
\author{Xiaosen Yang}
\email{yangxs@ujs.edu.cn}
\affiliation{\JSU}
\author{Dong Wang}
\email{dwang@ahu.edu.cn}
\affiliation{\ANU}
\date{\today}
\begin{abstract}

Environmental decoherence occurs when a quantum system interacts with its surroundings, progressively reducing quantum interference and coherence, complicating the preservation of critical quantum features over time, especially during experimental implementation. The quantum features of a state can be represented in phase space via the Wigner function, which manifests across multiple scales, with decoherence potentially influencing each scale differently, as examined in this work. We consider the compass state and its photon-added and photon-subtracted variants (optimized compass states) as our representative examples, each of which exhibits phase-space features with dimensions beyond the Planck scale, making them suitable for quantum sensing applications. We investigate the interaction of these states with a heat reservoir by employing a range of well-established theoretical tools. We observe that compass states with finer-scale phase-space features are more fragile to decoherence, with parameters favoring greater sub-Planckness in phase space concomitantly increasing the fragility of these compass states to decoherence. Our findings are then validated for generic quantum states interacting with the heat reservoir, for which we provide analytical and numerical investigations, exploring the relationship between quantum state robustness to decoherence and the sizes of their phase-space features; that is, phase-space features at smaller scales decay faster under decoherence, and vice versa.
\end{abstract}
\maketitle
\section{INTRODUCTION}

The phenomenon of interference is a crucial aspect of quantum mechanics~\cite{Gerry05book,ficek2005quantum}, capable of inducing nonclassical features between the composite components of a state~\cite{PhysRevX.9.021023,nemoto2003quantummetrology,pan2014,mitchell2004,Kwon2019,Zubairy2020,Jeong2001,Florian2018, Fabre2020, Mil86, YS86,Liu2024,Kienzler2016}, which can be characterized in position–momentum phase space via the Wigner function~\cite{Sch01,Davis2021,Chabaud2021,PhysRevX.5.021003}. Quantum interference in phase space can yield superoscillatory patterns, in which the phase space area of a fundamental structure may surpass the typical uncertainty principle constraints~\cite{Robertson1929}. This calls into question the widely held belief that phase-space features cannot be finer than the Planck scale ($\hbar$), demonstrating that quantum interference can develop phase-space features at sub-Planck scales ($\ll\hbar$)~\cite{Zurek2001,Howard2019}. Sub-Planck features are critical since they may play a significant role in quantum metrology~\cite{zheng2025,Eff4,Toscano06,deng2024quantum}.

Compass states~\cite{Zurek2001,Howard2019,naeem2021, naeem2022,PhysRevA.78.034101}, widely regarded as delivering sub-Planck features, can be understood as part of the broader class of circular superposition states~\cite{llhr-hn2y,hailin2024,8yqc-ynp4,PhysRevA.48.2213,PhysRevA.93.062323}. Multiphoton manipulations are crucial for advancing the nonclassical characteristics of quantum states~\cite{Chen2024, Wenger, Barnett2018, Guerrini, Zavatta_2008, Lund2024, Takase2021,naeem2023}. Their compatibility with modern experimental platforms provides a practical and accessible approach for generating nonclassical states~\cite{Alexei2006, Neergaard2006}. The application of photon addition and subtraction operations to the standard compass state (a superposition of four coherent states) results in the refinement of the corresponding phase space features~\cite{akhtar2024sub-shot}. This optimized version of the compass state may attain isotropic sub-Planck structures.

Decoherence is a major obstacle to preserving quantum
features of a state~\cite{Brody2025,jcd6lft3,PhysRevLett.86.2913,zurek_RevModPhys2003,gardiner2004, Akhtar2024,Hauer2023,Kolar2024negativewigner}, which become more challenging in the experimental settings~\cite{yu2025schrodinger,yang2024,Alexei2006,Neergaard2006,Takahashi2008,Etesse2015,Huang2015,lo2015spin,Kienzler2016}. In other words, decoherence is stressed as a fundamental barrier to the advancement of quantum information technologies~\cite{zurek_RevModPhys2003}, though its impact can be mitigated using a range of established techniques~\cite{Vitali1998,Jeannic2018,Pan2023,PhysRevD.100.016020,Liu_PhysRevApplied,Teh2020}.

In the Wigner phase space representation~\cite{Sch01}, nonclassical effects manifest as structures of differing scales, which can exhibit distinct susceptibility to environmental decoherence. In this work, we aim to identify the variation of the impact of decoherence with respect to the phase-space scales. For this, we start with two physically relevant examples exhibiting sub-Planck traits, the compass state~\cite{Zurek2001} and its optimized version~\cite{akhtar2024sub-shot}, under the impact of environmental decoherence. We employ the heat reservoir as a controlled source of decoherence~\cite{SCHLOSSHAUER20191, PhysRevA.83.062121}, which enables a simplified investigation to reach our goals. The interaction of the outlined compass states with the heat reservoir leads them to lose their quantum features and steers them toward classical mixtures, thus illustrating the quantum-to-classical transition. 

We demonstrate our findings through a series of theoretical investigations, analyzing the impact of decoherence on corresponding compass states. First, we present the Wigner functions and their temporal evolution after interaction with the heat reservoir, which shows that the quantum features, manifested as negative regions in the phase space~\cite{AnatoleKenfack2004}, gradually diminish, as indicated by the attenuation of corresponding Wigner negativity and sub-Planck structures. This degradation of quantum features is further emphasized in corresponding optical tomogram distributions, which provide additional insight by precisely highlighting the degraded regions of the Wigner distributions across specific directions in phase space~\cite{Leonhardt1995,Leonhardt1996,Vogel1989,Strandberg2022}. 
We additionally evaluate the loss of quantum coherence by discussing quantum purity during thermal contact of considered compass states~\cite{PhysRevA.83.062121}. We notice that either increasing the number of added photons or raising the amplitudes of the component coherent states in our superpositions favors the creation of sub-Planck structures. Although such finer-scale features are useful for quantum sensing, they may increase the susceptibility of these states to decoherence. In contrast, applying photon subtraction following photon addition reduces the fineness of these sub-Planck features, making the states more stable against thermal decoherence.

Subsequently, we validate our findings by evaluating the robustness of a generic quantum state interacting with the specified heat reservoir and analyze the impact of decoherence across varying sizes or extensions of phase-space features. The impact of decoherence on phase-space features of a quantum state appears to scale inversely with their sizes; tiny features (such as sub-Planck structures) decay more rapidly than larger ones. In other words, quantum features at smaller phase-space scales are more rapidly suppressed by environmental decoherence, indicating that they exhibit comparatively reduced stability.

This paper is organized as follows: We begin by characterizing the phase-space features of the representative compass states in Sec.~\ref{sec:cases}, followed by an evaluation of their robustness against decoherence using multiple theoretical frameworks in Sec.~\ref{sec:stability}. In Sec.~\ref{sec:phase_robustness1}, we develop a general framework for phase-space–scale-dependent robustness for any generic quantum state interacting with the heat reservoir. The results are summarized in Sec.~\ref{sec:conc}.

\begin{figure*}[t]
\centering
\includegraphics[width=\textwidth]{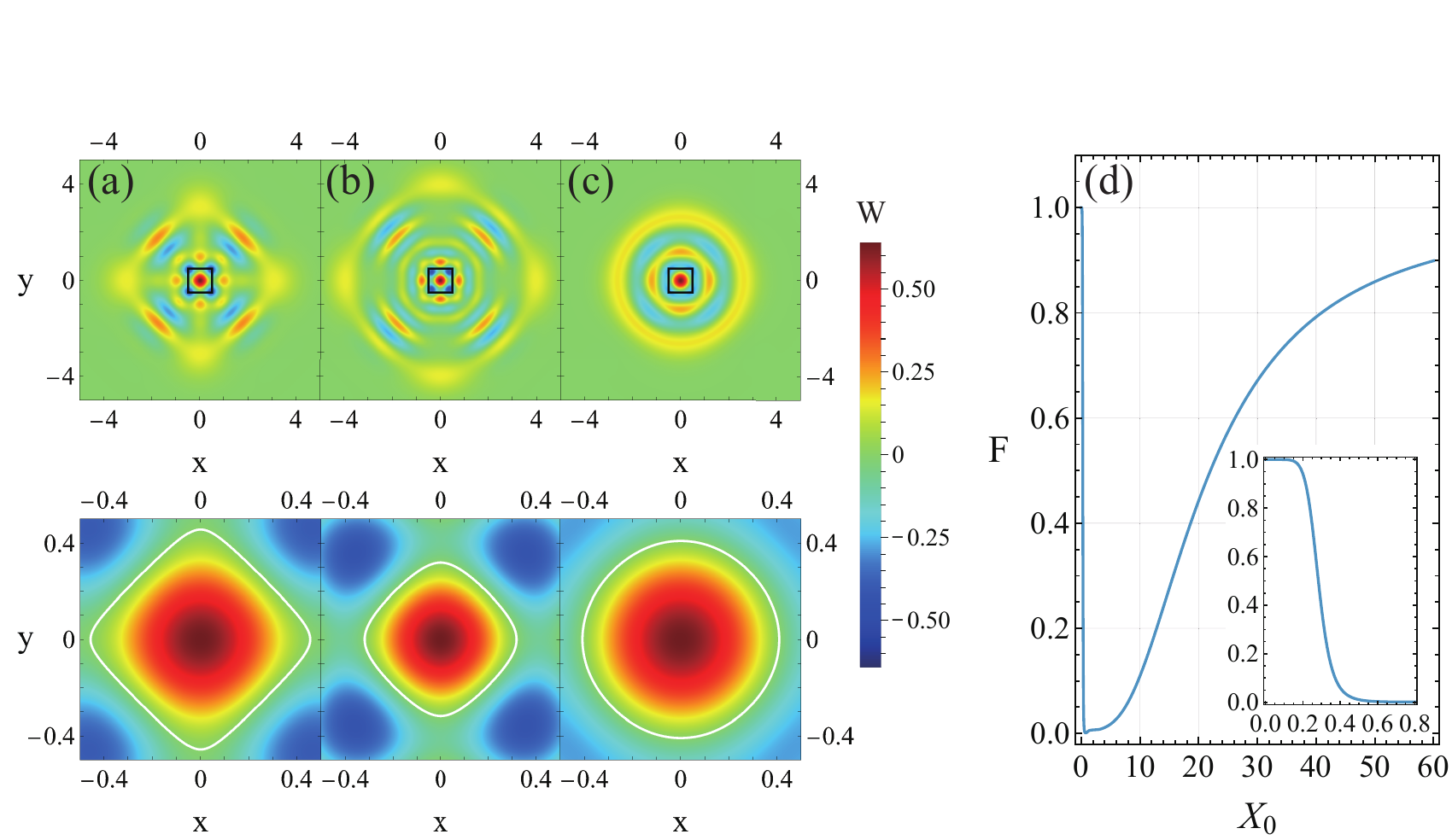}
\caption{The Wigner distributions $W_{\ket{\ltimes}}(\beta),$ with $\beta = \frac{(x+\mathrm{i}y)}{\sqrt{2}}$: (a) $p = q = 0$ and $X_0 = 3$; (b) $p = q = 14$ and $X_0 = 1.5$; (c) $p=q=14$ and $X_0=0.5$. The black box in (a)–(c) marks the focused phase space region, shown also across each main figure. The white contours represent regions where the Wigner distributions have virtually zero amplitudes near the origin. (d) The overlap $F = |\langle C | \ltimes \rangle|^2$ between the state $\ket{C}$ and its optimized version $\ket{\ltimes}$ in the case of $p = q = 14$.}
\label{fig:fig1}
\end{figure*}

\section{REPRESENTATIVE STATES AND SUBLEVELS} \label{sec:cases}

In this section, we review compass states and their related quantum states, specifically those obtained by applying multiphoton operations on coherent states and their superpositions. Compass states and their variants may exhibit sub-Planck features that enable sensitivity to displacements beyond the standard quantum limit and are regarded as valuable resources for quantum metrology~\cite{llhr-hn2y,akhtar2024sub-shot, Howard2019, Eff1, Eff3, naeem2021, naeem2022, Luo_2024, hailin2024}. The optimized compass state~\cite{akhtar2024sub-shot}, which is the central focus of this work, exemplifies these sub-Planck features and is incorporated into our discussion. Here, we mainly focus on the phase-space features of the optimized compass state~\cite{akhtar2024sub-shot}, comparing them with those of the standard compass state~\cite{Zurek2001}, and then we discuss Wigner negativity of these compass states.

\subsection{Phase space}

The Wigner function~\cite{Sch01}, a tool for describing quantum states in phase space, describes nonclassical structures in phase space. The nonclassicality of a state can be defined by the negativity of its Wigner distribution~\cite{PhysRevLett.122.080402,Davis2021, Chabaud2021, PhysRevX.5.021003}, while its marginals correspond to the statistics of quadrature operators~\cite{naeem2019}. Mathematically, the Wigner function for the density operator $\hat{\rho}$ is the expectation value of the parity kernel~\cite{Russel2021,Barnett2018} $W_{\hat{\rho}}(\beta)=\mathrm{tr}\big(\hat{\rho}\hat{\Delta}(\beta)\big)$. Alternatively, one may use the following mathematical form, which we employ in the present work~\cite{Sch01}
\begin{align}\label{eq:wigner_general}
W_{\hat{\rho}}(\beta)= \frac{2\mathrm{e}^{2|\beta|^2}}{\pi^2}\int \mathrm{d}^2\zeta\,\braket{-\zeta|\hat{\rho}|\zeta}\,\mathrm{e}^{-2\left(\beta^*\zeta-\beta\zeta^*\right)},
\end{align}
where $\beta=\frac{x+\mathrm{i}y}{\sqrt{2}}$, with $x$ and $y$ represent the position and momentum pairs, respectively. Wigner negativity is a basic measure of the nonclassicality of a quantum state~\cite{AnatoleKenfack2004}, and it can be stated mathematically as:

\begin{align}\label{eq:neg_vol}
   \delta=\int |W_{\hat{\rho}}(\beta)|\mathrm{d}^2\beta-1.
\end{align}
When the Wigner function $W_{\hat{\rho}}$ is non-negative, the quantity $\delta$ is equal to zero. This is evident in the case of coherent states and squeezed vacuum states, where $\delta = 0$~\cite{AnatoleKenfack2004}. Since squeezed vacuum states are recognized as nonclassical, this suggests that the parameter ($\delta$) alone may not be sufficient to fully characterize nonclassicality in all cases. However, the presence of negativity in the Wigner function offers enough evidence of nonclassicality in a state~\cite{AnatoleKenfack2004,Davis2021,Chabaud2021,PhysRevX.5.021003}.

Adding and subtracting photons from a quantum state may affect its nonclassical characteristics. Photon addition and subtraction, when applied with the appropriate selection and sequencing, can yield a nonclassical variant of the coherent state ($\ket{\alpha}$)~\cite{akhtar2024sub-shot}, denoted as $\ket{\Omega_{\alpha}}=\hat{a}^{q}\hat{a}^{\dagger p}\ket{\alpha}$ with $p$ and $q$ representing the quantity of added and subtracted photons, respectively. The Wigner function of the coherent state ($\ket{\alpha}$) exhibits a Gaussian peak and serves as a classical reference, while its photon-altered counterpart can exhibit negativity in the Wigner distribution, indicating its nonclassicality~\cite{AnatoleKenfack2004, Davis2021, Chabaud2021, PhysRevX.5.021003}. This negativity in the Wigner distribution is controlled by the multiphoton operations applied to the state. 

\subsection{Compass states}\label{subsec:cases}

Photon addition and subtraction applied to the compass state have resulted in an optimized version with improved phase-space characteristics~\cite{akhtar2024sub-shot}. The mathematical form of this optimized compass state is
\begin{align}\label{eq:psa_com} 
\ket{\ltimes}=&\frac{1}{\sqrt{\aleph^\ltimes}}\left(\ket{X}+\ket{Y}\right)
\end{align}
with $\ket{X}=\ket{\Omega_{\alpha_1}}+\ket{\Omega_{\alpha_2}}$ and $\ket{Y}=\ket{\Omega_{\alpha_3}}+\ket{\Omega_{\alpha_4}}$, where $\alpha_1=\nicefrac{X_0}{\sqrt{2}}$, $\alpha_2=\nicefrac{-X_0}{\sqrt{2}}$, $\alpha_3=\nicefrac{\text{i}X_0}{\sqrt{2}}$ and $\alpha_4=\nicefrac{-\text{i}X_0}{\sqrt{2}}$ with $X_0\in  \mathbb{R}^+$. The normalization coefficient $\aleph^ \ltimes$ is denoted as
\begin{align}
\aleph^\ltimes=&\nonumber (-1)^{p+q}\sum^4_{i,j=1}\sum^p_{n=0}\frac{(-1)^n (p!)^2}{n![(p-n)!]^2}G_{\alpha_i,\alpha_j} \exp(\alpha^*_i\alpha_j)\\&\times H_{p-n,q}\left(\text{i}\alpha_j,\text{i}\alpha^*_i\right) H_{p-n,q}\left(\text{i}\alpha^*_i,\text{i}\alpha_j\right)
\label{eq:norm1}
\end{align}
with
\begin{align}
G_{\alpha_i,\alpha_j}=\exp\left[-\frac{1}{2}\left(|\alpha_i|^2+|\alpha_j|^2\right)\right].
\end{align}
The generating function of the bivariate Hermite polynomial $H_{m,n}(.)$ is given by~\cite{meng2020multi}:
\begin{equation}\label{hermite}
    H_{m,n}(x_1,y_1)=\frac{\partial^{n+m}}{\partial s^n \partial t ^m}\text{e}^{-s t+s x_1+t y_1}\big|_{s,t=0},
\end{equation}
which is heavily used in our calculations.

The optimized compass state presented in Eq.~(\ref{eq:psa_com}) can be reduced to different variants under suitable choices of $p$, $q$, and $X_0$. In the absence of photon addition and subtraction ($p=0$ and $q=0$), the resulting state reduces to a superposition of four coherent states, or equivalently, a superposition of two ordinary cat states, denoted as
\begin{align}\label{eq:zurek_compass}
\ket{C}=\frac{1}{\sqrt{N_c}}\left[\Ket{\frac{X_0}{\sqrt{2}}}+\Ket{\frac{-X_0}{\sqrt{2}}}+\Ket{\frac{\text{i}X_0}{\sqrt{2}}}+\Ket{\frac{-\text{i}X_0}{\sqrt{2}}}\right],
\end{align}
where the normalization coefficient ($N_c$) can be obtained by setting $p=q=0$ in Eq.~(\ref{eq:norm1}). When the four components of the state $\ket{C}$ are sufficiently distinguishable, i.e., when pairs with the same amplitudes but opposite phases form cat states such that $\Braket{\frac{X_0}{\sqrt{2}}|\frac{-X_0}{\sqrt{2}}}\approx 0$ and $\Braket{\frac{\mathrm{i}X_0}{\sqrt{2}}|\frac{-\mathrm{i}X_0}{\sqrt{2}}}\approx 0$ are satisfied for $X_0 \gg 1$, the state corresponds to a superposition of two macroscopic cat states, yielding the well-known compass state~\cite{Zurek2001}, which otherwise can be considered as the superposition of two kitten states. Other variants, with appropriate choices of photon operations, can also exhibit crucial phase-space features~\cite{akhtar2024sub-shot}.

The Wigner function of the optimized compass state $\ket{\ltimes}$ can be obtained using Eq.~(\ref{eq:wigner_general}), yielding the following compact analytical expression:
\begin{align}\label{eq:wig_psa}
    W_{\ket\ltimes}(\beta)=\frac{1}{\aleph^{\ltimes}} \left(\sum^4_{i=1}W_{\ket{\ltimes_i}\bra{\ltimes_i}}(\beta)+\sum_{i\ne j=1}^{4}W_{\ket{\ltimes_i}\bra{\ltimes_j}}(\beta)\right),
\end{align}
where the first term corresponds to the photon-added and then photon-subtracted coherent states, while the second term corresponds to the nonclassical effects associated with these states, and
\begin{align}\label{eq:psa_wig}
 W_{\ket{\ltimes_i}\bra{\ltimes_j}}(\beta)=&\nonumber W_{\ket{\alpha_i}\bra{\alpha_j}}(\beta)\sum^p_{n=0} \frac{(-1)^n (p!)^2}{n![(p-n)!]^2} \\&\times H_{p-n,q}\left(\text{i}\gamma^*_j,\text{i}\alpha_i\right) H_{p-n,q}\left(-\text{i}\gamma_i,-\text{i}\alpha^*_j\right)
\end{align}
with
\begin{align}
    \gamma_k=2\beta-\alpha_k,
\end{align}
\begin{align}\label{eq:gen1_wig}
W_{\ket{\alpha_i}\bra{\alpha_j}}(\beta)=&\nonumber G_{\alpha_i,\alpha_j}\exp\Big[-\alpha_i\alpha^*_j-2\big(|\beta|^2-\alpha^*_j\beta\\ &-\alpha_i\beta^*\big)\Big].
\end{align}
Figure~\ref{fig:fig1} depicts heat maps of the Wigner functions of the compass state and its optimized variant, illustrating the variations in associated phase-space features by varying the parameters $X_0$, $p$, and $q$. The central region emphasized in the black box [Fig.~\ref{fig:fig1} (a)-(c)] represents the central sub-Planck structure, as also separately illustrated for each case, with white contours representing the zeros of the Wigner distribution around the phase space origin. These contours highlight the borders of the fundamental sub-Planck structure and aid in distinguishing between its isotropic and anisotropic forms.

\begin{figure}[t]
\centering
\includegraphics[width=0.45\textwidth]{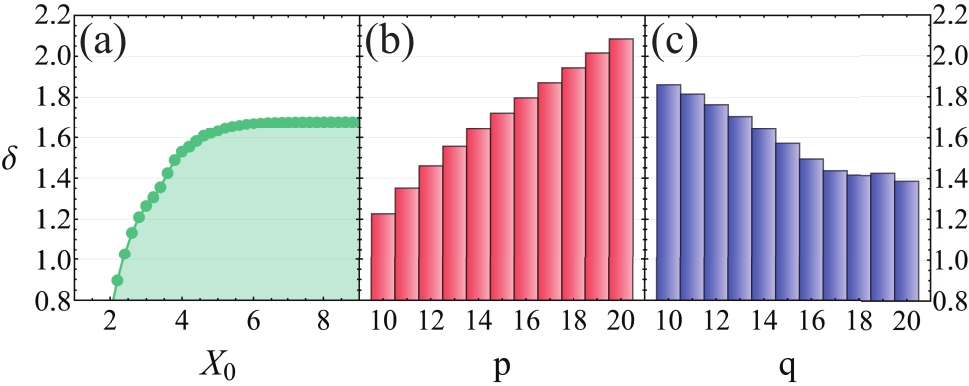}
\caption{Variation of negativity in the Wigner function $W_{\ket{\ltimes}}(\beta)$: (a) $p=q=0$, (b) $q=14$, $X_0=1.5$, and (c) $p=14$, with $X_0=1.5$.}
\label{fig:fig2}
\end{figure}

The Wigner function of the standard compass state is illustrated in Fig.~\ref{fig:fig1}(a), where the central sub-Planck structure is confined within an anisotropic region (a tilelike pattern), clearly demonstrating its directional asymmetry. Nonzero values of $p$ and $q$ combined with a relatively small $X_0$ give rise to an optimized version of the compass state, in which variations in $X_0$ modify the central sub-Planck structure as depicted in Fig.~\ref{fig:fig1}(b) when $p = q = 14$ and a relatively smaller $X_0$, evidencing similar phase-space characteristics to those in Fig.~\ref{fig:fig1}(a), including a central anisotropic sub-Planck structure. A further decrease in the value of $X_0$, while keeping the same $p$ and $q$ values as in Fig.~\ref{fig:fig1}(b), results in the optimized compass state displaying an isotropic sub-Planck structure, as shown in Fig.~\ref{fig:fig1}(c). This isotropy is characterized by the confinement of the feature within a circular region. This isotropic form is a superior manifestation of sub-Planck traits, as it enables uniform sensitivity enhancement to phase-space displacements in all directions~\cite{hailin2024}. Notably, the optimized compass state considered here was overlooked in earlier work~\cite{akhtar2024sub-shot} regarding its isotropic sub-Planck features, in contrast to states produced by reversing photon operations (e.g., subtraction followed by addition). We demonstrate that an appropriate choice of the coherent-state amplitudes in the superposition can likewise produce isotropic sub-Planck structures, and the emergence of sub-Planck features in these cases can be attributed to multiphoton operations. Similar isotropic sub-Planck features are also contained in multicomponent cat states~\cite{hailin2024}, composed of superpositions of more than four coherent states. However, the optimized compass state obtained via photon addition followed by subtraction, as well as its reverse-order variant~\cite{akhtar2024sub-shot}, is distinct from multicomponent cat states, demonstrating that such photon operations on a standard compass state can also produce isotropic sub-Planck features.

The transition between the compass state and its optimized counterpart is solely determined by the choice of parameters $X_0$, $p$, and $q$. Smaller values of the parameter $X_0$ provide a higher distinguishability between the standard compass state and its optimized version, as evident in Fig.~\ref{fig:fig1}(d), where we plot the overlap $F(X_0) = |\langle C | \ltimes \rangle|^2$, which quantitatively compares the compass state with its multiphoton-altered version. This shows that the photon addition followed by subtraction significantly affects the state with a particular choice of $X_0$. For smaller values of $X_0$, the optimized version of the compass state differs notably from the standard compass state, as also confirmed by their Wigner distributions, and in these situations the photon operations applied are highly effective in deforming the state. When the parameter $X_0$ increases, the effectiveness of photon operations diminishes, and at relatively large $X_0$ values, the optimized compass state gradually converges toward its original form~\cite{Zurek2001}.

For the optimized compass state~\cite{akhtar2024sub-shot}, specific choices of photon addition and subtraction have a pronounced effect on the critical sub-Planck features. As discussed~\cite{akhtar2024sub-shot}, increasing the number of added photons generally favors the emergence of sub-Planck features, whereas photon subtraction alone does not affect the sub-Planckian characteristics. Note that the compass state is an eigenstate of $\hat{a}^4$, indicating that photon subtraction multiple of 4 leaves the state essentially unchanged. However, when combined with photon addition, photon subtraction tends to compress the core sub-Planck structures, while photon addition has the opposite effect, enhancing their resolution. To induce sub-Planck features in the optimized version of the compass state, a sufficient number of photon additions and subtractions is required, governing the formation and suppression of these features, respectively.

We further investigate the effect of multiphoton operations on the nonclassical characteristics of the optimized compass state by evaluating its Wigner negativity using Eq.~(\ref{eq:neg_vol}). The results for various parameter choices are presented in Fig.~\ref{fig:fig2}. Figure~\ref{fig:fig2}(a) shows the case of the compass state ($p=q=0$), where we plot Wigner negativity ($\delta$) over a few $X_0$ values, which shows that the negative volume of the associated Wigner function increases as the amplitude of coherent states $X_0$ rises, up to a certain point, beyond which the negativity becomes constant and remains invariant for further increases in $X_0$. Figures \ref{fig:fig2}(b) and \ref{fig:fig2}(c) show the Wigner negativity of the optimized compass state for a fixed $X_0$, while the number of added photons ($p$) and subtracted photons ($q$) are varied. This demonstrates that, with an appropriate choice of order and application, photon addition ($p$) enhances the Wigner negativity of the optimized compass state, whereas photon subtraction ($q$) reduces it~\cite{akhtar2024sub-shot}. The Wigner negativity for other parameter choices, including cases with intermediate photon addition ($p$) and subtraction ($q$) and different ranges of $X_0$, is analyzed in the Appendix, as shown in Fig.~\ref{fig1:appendix}. These results preserve the trends in Wigner negativity discussed in this section.

\section{PHASE SPACE AND DECOHERENCE: SPECIFIC CASES}\label{sec:stability}
As discussed in Sec.~\ref{sec:cases}, the crucial phase-space features of the standard compass state and its variant are influenced by different parameters, with the standard compass state representing a limiting case of the optimized one; that is, by adjusting certain parameters, these two states can be easily transformed into each other. In this section, we present the main results of our investigation by considering the compass state and its optimized versions, examining how the effects of decoherence evolve during the transition between these states through their interaction with a thermal reservoir~\cite{Kim1992, PhysRevA.63.032106}.

The master equation governing the temporal evolution of a single-mode state $\hat{\rho}$ in the presence of the considered thermal reservoir is given by:
\begin{align}\label{eq:models_decoh}
 \frac{\partial}{\partial t} \hat{\rho}=&\nonumber\omega\left(\overline{n}+1\right)(2\hat{a}\hat{\rho}\hat{a}^{\dagger}-\hat{a}^{\dagger}\hat{a}\hat{\rho}-\hat{\rho}\hat{a}^{\dagger}\hat{a})\\&+\omega \overline{n}(2\hat{a}^{\dagger}\hat{\rho}\hat{a}-\hat{a}\hat{a}^{\dagger}\hat{\rho}-\hat{\rho}\hat{a}\hat{a}^{\dagger}),
\end{align}
where the first term represents the transfer through the decay of photons from the quantum system to the thermal reservoir, and the second term corresponds to the transfer of excitation from the non-zero temperature heat reservoir to the quantum system. The field decay rate is labeled by $\omega$, and the average thermal photon number in the cavity is $\overline{n}$. When $\overline{n}=0$, Eq.~(\ref{eq:models_decoh}) depicts a zero-temperature reservoir, which is also known as the photon-loss reservoir~\cite{Buzek1992}. 

In this section, we mainly analyze the impact of decoherence on the phase-space features of the compass state and its optimal version introduced in Sec.~\ref{subsec:cases}, employing a range of theoretical methods. The discussion is detailed in the following sections. Sec.~\ref{subsec:Wig_func} examines the time evolution of the Wigner distributions, while Sec.~\ref{subsec:tomo} analyzes the tomogram distributions of the considered compass states. The stability of the sub-Planck structures under decoherence is investigated in Sec.~\ref{subsec:sub_Planck_decay}, and the temporal evolution of the resulting Wigner negativity is discussed in Sec.~\ref{subsec:neg_decay}. Finally, Sec.~\ref{subsec:entropy} presents the time evolution of the linear entropy.

\begin{figure}[htp!]
\centering
\includegraphics[width=0.53\textwidth]{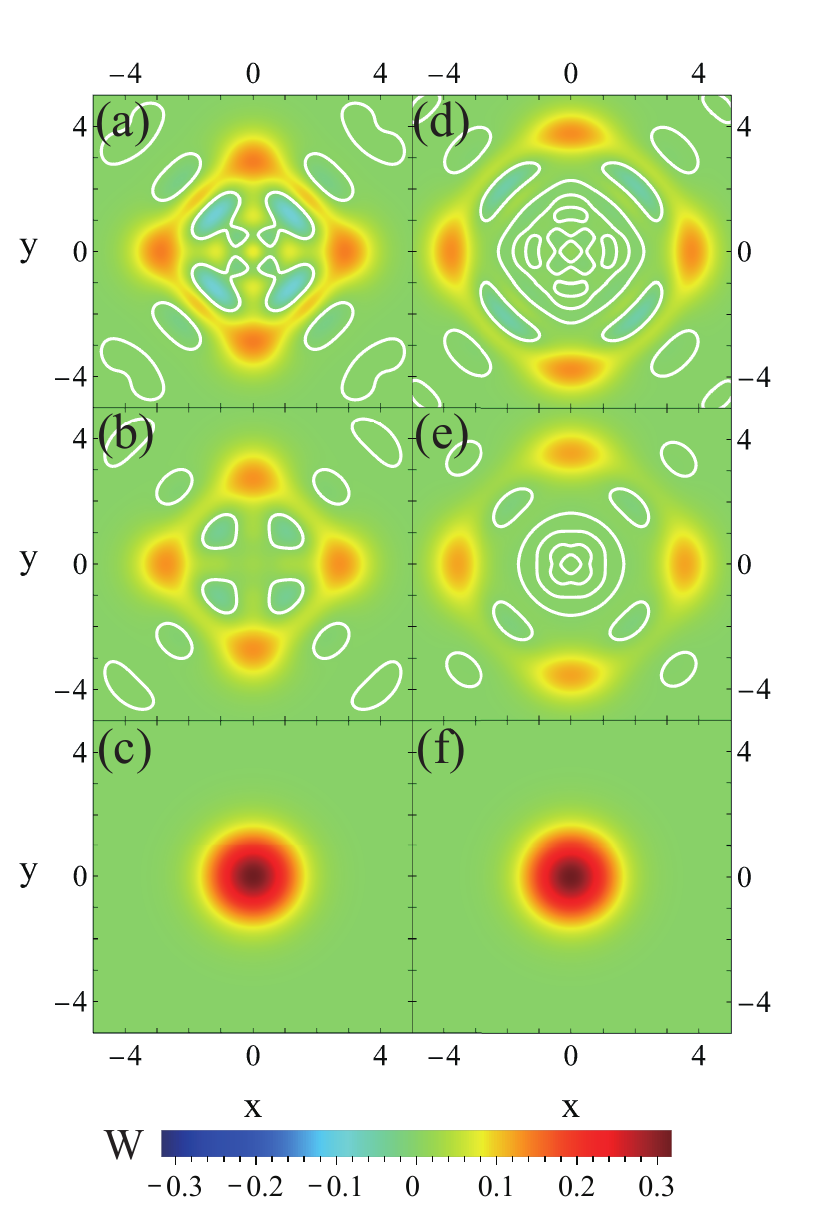}
\caption{The time evolution of the Wigner functions corresponding to our representative states. The white contour highlights the regions of the Wigner distribution that have substantially decayed under decoherence, ultimately evolving toward a Gaussian lobe (a thermal state) over time. The left column represents the cases when $p = q = 0$, $X_0 = 3$, and the right column belongs to $p = q = 14$, $X_0 = 1.5$, each evaluated at multiple values of $\tau$. (a)~$\tau = 0.06$, (b)~$\tau = 0.12$, (c)~$\tau = 10$, (d)~$\tau = 0.06$, (e)~$\tau = 0.12$, and (f)~$\tau = 10$. In all cases, we set $\overline{n} = 0.5$.}
\label{fig:fig3}
\end{figure}

\subsection{Damping of interference fringes}\label{subsec:Wig_func}
This section focuses on the temporal evolution of the Wigner function of the optimized compass state [Eq.~(\ref{eq:psa_com})] under the influence of the thermal reservoir described by Eq.~(\ref{eq:models_decoh}), considering a range of parameter settings. The standard compass state is treated as a special case within the general framework of the optimized compass state. The temporal evolution of the Wigner distribution associated with the optimized compass state under the influence of the reservoir is computed using~\cite{hu_time_2009} 
\begin{equation}\label{eq:solution}
 W_{\hat{\rho}}(\eta,t)=\frac{2}{\midpoint{T}}\int\frac{\mathrm{d}^2\beta}{\pi} W_{\hat{\rho}}(\beta)\exp\left[\frac{-2\left|\eta-\beta\text{e}^{-\omega t}\right|^2}{\midpoint{T}}\right],   
\end{equation}
where $\eta=\frac{(x + \text{i}y)}{\sqrt{2}}$ and $\midpoint{T}=(1 + 2 \overline{n})T$, with $T=1-\text{e}^{-2\omega t}$.

Since the optimal compass state is constructed by employing the sequence of annihilation and creation operators on the standard compass state, which in this case are in the antinormal order ($\hat{a}$ is placed before $\hat{a}^\dagger$), which makes the calculations leading to the temporal evolution of the resulting Wigner functions highly complicated, and rewriting these operators in their normal order as $\hat{a}^{n}\hat{a}^{\dagger m}=(-\text{i})^{n+m}: H_{n,m}(\text{i} \hat{a}^{\dagger},a):$ makes the calculation easier~\cite{Hong-Yi_2002}, but the presence of the bivariate Hermite polynomial $H_{n,m}$ in the integral still complicates the calculation, which using the generating function as provided in Eq.~(\ref{hermite}) makes it comparatively easier to reach compact analytical expressions.

We also make frequent use of the identity given by,
\begin{equation}
\exp(Cs+Dt+Est)
=\sum_{l=0}^{\infty}\frac{E^{l}}{l!}\frac{\partial^{2l}}{\partial C^{l}\partial D^{l}}\left[\exp\left(Cs+Dt\right)\right],
\end{equation}
and the following integral formula makes our derivations relatively simple~\cite{puri2001}
\begin{align}\label{eq:int_1}
        &\nonumber\int^{\infty}_{-\infty}\mathrm{d}^2 \beta \exp{\bigg[A |\beta|^2+B \beta+C \beta^*+D\beta^2+K\beta^{* 2}}\bigg]\\=&\frac{\pi}{\sqrt{A^2-4 D K}}\exp{\bigg[\frac{-A B C+B^2 K+C^2 D}{A^2-4 D K}}\bigg],
    \end{align}
whose convergent conditions are $\operatorname{Re}\left[A\pm D\pm K\right]<0$ and $\operatorname{Re}\left[\frac{(A^2-4 D K)}{A\pm D \pm K}\right]<0$. 

Finally, the temporal evolution of the Wigner function belonging to the optimal compass state in the designated heat reservoir is expressed in the following compact mathematical form:
\begin{align}\label{eq:tempo_psa}
    W_{\ket\ltimes}(\eta,\tau)=\frac{1}{\aleph^{\ltimes}} \Big[\text{I}_{\text{coh}}(\eta,\tau)+\text{I}_{\text{int}}(\eta,\tau)\Big]
\end{align}
with $\text{I}_{\text{coh}}(\eta,\tau)=\sum^4_{i=1}W_{\ket{\ltimes_i}\bra{\ltimes_i}}(\eta,\tau)$ and 
$\text{I}_{\text{int}}(\eta,\tau)=\sum_{i\ne j=1}^{4}W_{\ket{\ltimes_i}\bra{\ltimes_j}}(\eta,\tau)$, where
\begin{align}
    W_{\ket{\ltimes_i}\bra{\ltimes_j}}(\eta,\tau)=&\nonumber\frac{2 \exp \big[A-\frac{2|\eta|^2}{\midpoint{T}}-\alpha_i \alpha^*_j\big]G_{\alpha_i,\alpha_j}}{(k_3+2)\midpoint{T}}\\&\nonumber\times \sum^p_{n=0}\sum^{\infty}_{n^\prime=0}\frac{(-1)^n(p!)^2 (E)^{n^\prime}}{n! n^\prime ![(p-n-n^\prime)!]^2}\\&\times H_{p-n-n^\prime,q}[\Gamma_j,\text{i}\alpha_i]H_{p-n-n^\prime,q}[\Gamma_i,-\text{i}\alpha_j]
\end{align}
with
\begin{align}
  &E=\frac{4}{(2+k_3)},A=\frac{(k_1+2\alpha_i)(k_2+2\alpha^*_j)}{(2+k_3)},\\&\Gamma_j=\frac{2\text{i}(k_2+2\alpha^*_j)}{(2+k_3)}-\text{i}\alpha^*_j,\Gamma_i=\frac{-2\text{i}(k_1+2\alpha_i)}{(2+k_3)}+\text{i}\alpha_i,
\end{align}
where $k_1=\frac{2\text{e}^{-\tau}\eta}{\midpoint{T}},~k_2=\frac{2\text{e}^{-\tau}\eta^*}{\midpoint{T}}$, $k_3=\frac{2\text{e}^{-2\tau}}{\midpoint{T}}$, and $\tau=\omega t$.

The time evolution of the Wigner function associating the optimized compass state, as provided in Eq.~(\ref{eq:tempo_psa}), is now analyzed under the influence of the heat reservoir by varying the time parameter $\tau$, while keeping the average thermal photon number $\overline{n}$ fixed and tuning other relevant parameters. We begin with the case where $p = q = 0$ and $X_0 = 3$, which simply corresponds to the standard compass state~\cite{Zurek2001}, and the temporal evolution of the associated Wigner function under the influence of the heat reservoir is shown in Figs.~\ref{fig:fig3}(a)–\ref{fig:fig3}(c) for different values of the time parameter $\tau$. Specifically, Fig.~\ref{fig:fig3}(a) illustrates the case at a relatively short time ($\tau = 0.06$), indicating that even after such a brief interaction with the heat reservoir, the compass state has lost a substantial amount of its nonclassical features.

The degradation of nonclassical features is clearly evident, with the white contour indicating the regions where the Wigner distribution has been degraded. This highlights that the central sub-Planck features quickly blur due to decoherence, illustrating the negative impact of decoherence on the crucial features of the state. Figure~\ref{fig:fig3}(b) indicates that a longer interaction with the heat reservoir ($\tau = 0.12$) leads to further degradation of the nonclassical phase-space features of the state. As illustrated in Fig.~\ref{fig:fig3}(c), when the interaction becomes significantly longer ($\tau = 10$), the nonclassical features of the state are completely lost, and the Wigner function emerges as a Gaussian lobe. When the amplitude $X_0$ is reduced, multiphoton operations become significantly more effective, resulting in the formation of optimized compass states. In the illustrated cases [Figs.~\ref{fig:fig3}(d)-\ref{fig:fig3}(f)], we uniformly set $p = q = 14$ and $X_0=1.5$, and the analysis is carried out over different time intervals: $\tau = 0.06$ in Fig.~\ref{fig:fig3}(d), $\tau = 0.12$ in Fig.~\ref{fig:fig3}(e), and $\tau = 10$ in Fig.~\ref{fig:fig3}(f). As the time $\tau$ rises, the optimized compass state progressively loses its nonclassical characteristics, where the faded regions are indicated by the white contours.

For sufficiently large values of $\tau$, the Wigner distribution transforms into a Gaussian lobe, resembling that of the standard compass state. This indicates that at higher $\tau$, these states transition into a thermal state, characterized by the Wigner function ($W_\text{th}(\eta) = [\pi (2\overline{n}+1)]^{-1} \exp\left[-\frac{2|\eta|^2}{2\overline{n}+1}\right]$). This indicates a quantum-to-classical transition, where the system reaches thermal equilibrium, and the Wigner function of the resulting thermal state is solely influenced by the average thermal photon number $\overline{n}$, reducing to the vacuum state when $\overline{n} = 0$.

\subsection{Distorted tomogram distributions}\label{subsec:tomo}

The temporal evolution of the Wigner distributions evidenced that the interaction of our compass states with the heat reservoir has decreased associated nonclassical characteristics. In this section, the corresponding tomogram distribution highlights this degradation more apparently.

The quadrature histograms (tomogram function), denoted as $R(x_\theta)$, can be interpreted as marginal distributions. They correspond to the projections (Radon transform) of the Wigner function $W$ given by~\cite{Leonhardt1995}
\begin{align}\label{eq:histro}
R(x_\theta)=\int \mathrm{d}y_\theta W(x_\theta \cos \theta-y_\theta \sin \theta,x_\theta \sin \theta+y_\theta \cos \theta).
\end{align}
The inverse Radon transform can be employed to reconstruct the Wigner function from these histograms $R(x_\theta)$~\cite{Vogel1989}. Note that in Eq.~(\ref{eq:histro}) $\theta$ corresponds to different directions of the phase space. We numerically computed Eq.~(\ref{eq:histro}) for our optimized compass state [Eq.~(\ref{eq:psa_com})] and investigated multiple scenarios under different parameter choices, which we now explain.

\begin{figure}[htp!]
\centering
\includegraphics[width=0.45\textwidth]{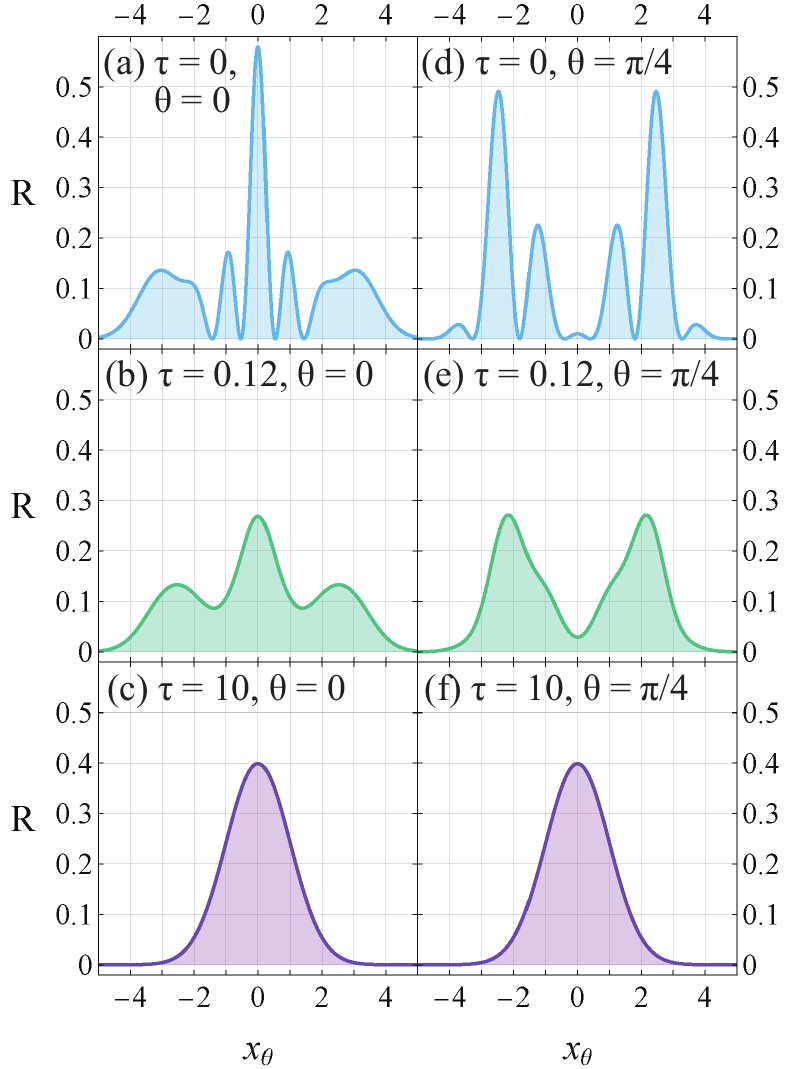}
\caption{Tomogram function for the case when $p = q = 0$ and $X_0=3$, illustrating the compass state.}
\label{fig:fig7}
\end{figure}

First, consider the tomogram distribution ($R$) for $p = q = 0$ and $X_0 = 3$, simply leading to the standard compass state~\cite{Zurek2001}. This case is illustrated in Fig.~\ref{fig:fig7} for different values of $\theta$ and $\tau$, where $\theta = 0$ corresponds to the $x$ direction in phase space, with the tomograms shown in Figs.~\ref{fig:fig7}(a)–\ref{fig:fig7}(c) associating with different values of $\tau$. Figure \ref{fig:fig7}(a) shows the case when $\tau=0$, representing the non-deformed form of the marginal distribution of the standard compass state along the $x$ direction, the oscillatory pattern around the origin mirrors the central interference pattern. When the interaction between the compass state and the heat reservoir is significant, the marginal distribution is distorted due to decoherence, as clearly observed in Figs.~\ref{fig:fig7}(b) and \ref{fig:fig7}(c) for $\tau = 0.12$ and $\tau = 10$, respectively. For a sufficiently large value of $\tau$ the Gaussian peak as shown in Fig.~\ref{fig:fig7}(c) evidence that the compass state has been transformed into a thermal state. $\theta=\frac{\pi}{4}$ represents diagonal direction in the phase space, which is observed for $\tau=0$, $\tau=0.12$ and $\tau=10$, and is illustrated in Figs.~\ref{fig:fig7}(d), \ref{fig:fig7}(e) and \ref{fig:fig7}(f), respectively. The deformation in these marginals is pronounced as the time $\tau$ increases, where for comparatively significant values of $\tau$, as observed in Fig.~\ref{fig:fig7}(e) corresponding marginal distribution is modified entirely compared to the prior case of Fig.~\ref{fig:fig7}(d), and then for sufficiently large values of $\tau$, as depicted in Fig.~\ref{fig:fig7}(f), a Gaussian peak emerges, reflecting the thermal state.

\begin{figure}[htp!]
\centering
\includegraphics[width=0.45\textwidth]{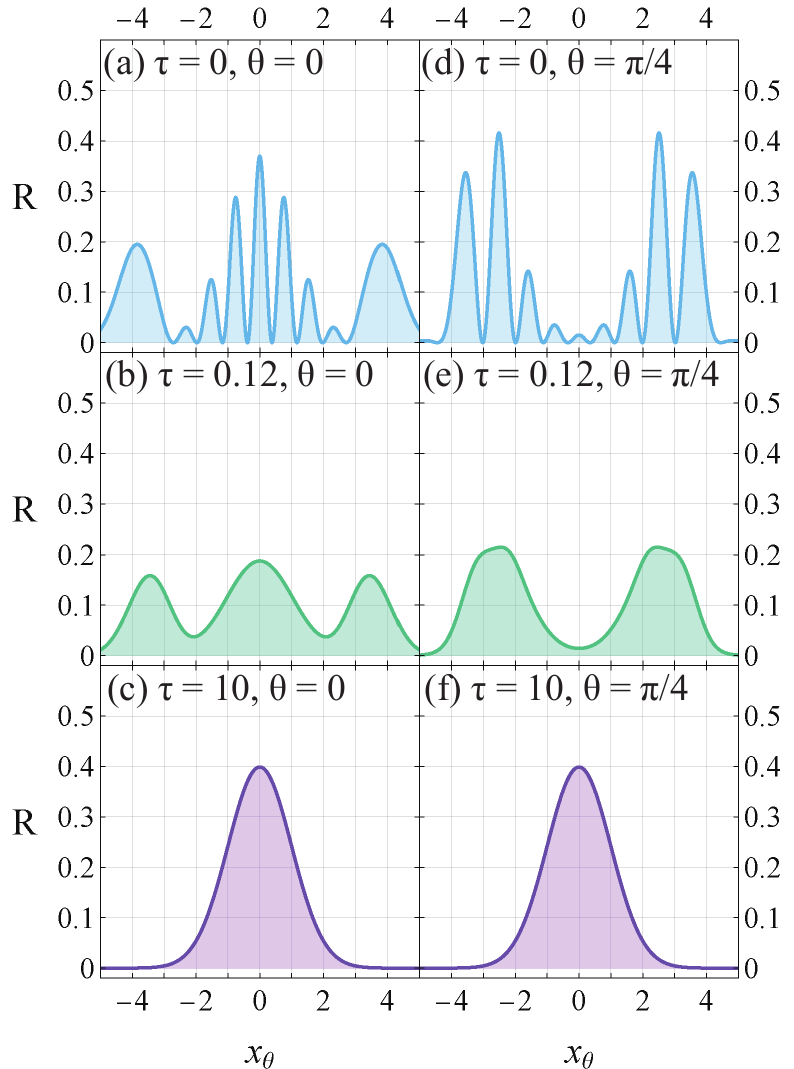}
\caption{Tomogram function for the case when $p = q = 14$ and $X_0=1.5$, illustrating the optimized compass state.}
\label{fig:fig8}
\end{figure}

We now discuss the cases of the optimized compass state. Figure~\ref{fig:fig8} represents the quadrature histograms ($R$) for the case when $p = q = 14$ and $X_0 = 1.5$, illustrated in different phase-space directions by varying the angle $\theta$; specifically, $\theta = 0$ corresponds to the $x$ direction in phase space and is further examined in Figs.~\ref{fig:fig8}(a)–\ref{fig:fig8}(c) for different values of $\tau$. Figure \ref{fig:fig8}(a) represents the situation when $\tau = 0$ denoting the non-distorted version of the marginal distribution. As time evolves, the associated marginal distribution becomes increasingly distorted. For example, at $\tau = 0.12$ [Fig.~\ref{fig:fig8}(b)], the distribution is notably altered from its initial shape. Eventually, at a much higher time ($\tau = 10$), as shown in Fig.~\ref{fig:fig8}(c), it fully transforms into a Gaussian profile.

We now expand our observations along the $\frac{\pi}{4}$ direction in the phase space, as illustrated in Figs.~\ref{fig:fig8}(d)–\ref{fig:fig8}(f), the increment in the values of $\tau$ has modified the associated marginal distributions. Initially, at $\tau = 0$, shown in Fig.~\ref{fig:fig8}(d), the marginal distribution takes its original form. As time progresses, the distribution begins to deform; for example, at $\tau = 0.12$, depicted in Fig.~\ref{fig:fig8}(e), the marginal distribution shows significant distortion. At sufficiently larger values of $\tau$, as shown in Fig.~\ref{fig:fig8}(f), a Gaussian peak emerges, marking the complete loss of associated quantum signatures.

In summary, thermal decoherence has diminished the critical nonclassical features of both the compass state and its optimized version. This loss is well apparent in the Wigner distributions, where Gaussian lobes emerge in phase space, and similarly reflected in the tomogram functions. The growing prominence of Gaussian peaks in these distributions is the message of the gradual transition from the quantum compass state to a thermal state.

\subsection{Fragility of sub-Planckness}\label{subsec:sub_Planck_decay}

\begin{figure}
\centering
\includegraphics[width=0.45\textwidth]{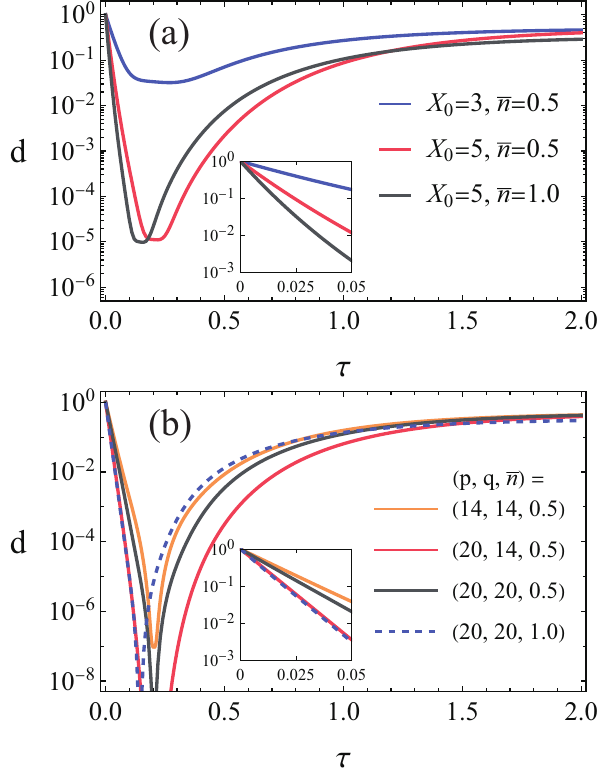}
\caption{The parametric function $d$ illustrates fluctuations in the height of the central phase-space features of the compass states under decoherence: (a)~$p = q = 0$; (b)~$X_0 = 1.5$.}
\label{fig:fig4}
\end{figure}

In this section, we investigate the impact of decoherence on the central sub-Planck structures of the outlined compass states. We examine the effects of main parameters, such as photon addition $p$, photon subtraction $q$, amplitude $X_0$, and the average number of thermal photons $\overline{n}$, on the prominence of these features using the following expression:
\begin{equation} 
d(\tau)= \left| \frac{W_{\ket\ltimes}(0,\tau)}{W_{\ket\ltimes}(0,0)} \right|,
\end{equation}
and using this equation, multiple scenarios relating to our quantum compass states are depicted in Fig.~\ref{fig:fig4}. In each of these depicted situations, a consistent pattern develops in the associated plots. The central sub-Planck structure diminishes gradually, and a new central peak emerges. This transition from the decay to the rebuilding of the central peak, with a clear distinction between the two phases, shows the degradation of quantum states as they shift from their quantum nature to a classical regime. The curve eventually levels off, signaling the complete decay of the core sub-Planck structure and the emergence of a new central feature.

The rate at which the central sub-Planck structure deforms into a new phase space feature is controlled by different associated parameters. This deformation is particularly noticeable in the shorter time frames, as shown in the inset of each case illustrated in Fig.~\ref{fig:fig4}. In Fig.~\ref{fig:fig4}(a), we explore scenarios where $p = q = 0$ (compass state) and analyze the effects of varying the parameters $X_0$ and $\overline{n}$. By comparing the blue and red curves in the inset of Fig.~\ref{fig:fig4}(a), it is observed that the central sub-Planck peak associating the compass state decays more quickly when the amplitude parameter $X_0$ grows. Similarly, the central peak decays faster with increasing average thermal photon number ($\overline{n}$) in the reservoir, as shown by the comparison of the red and black curves in the inset of Fig.~\ref{fig:fig4}(a).

We now discuss the scenario where the parameter $X_0$ is held constant while $p$ and $q$ varies (optimized compass state), as illustrated in Fig.~\ref{fig:fig4}(b). First, when $p$ is changed while keeping other parameters constant, as clearly illustrated in the inset of Fig.~\ref{fig:fig4}(b), the red curve falls more quickly than that of the orange curve, showing that larger photon addition may cause a rise in the decoherence effect. As the number of photon subtractions $q$ increases, as illustrated in the corresponding curves (compare the red and black curves) in the inset of Fig.~\ref{fig:fig4}(b), the black curve, which corresponds to a higher number of photon subtractions, decays more slowly than the red curve. This suggests that increasing the number of photon subtractions can help to stabilize the sub-Planck structures against decoherence. We further observe that larger average thermal photons $\overline{n}$ lead to increased decoherence effects on the sub-Planck structures of the optimized compass state, as observed by comparing the black and dotted blue curves of the inset shown within Fig.~\ref{fig:fig4}(b). Additional scenarios with comparatively smaller values of photon addition ($p$) and subtraction ($q$), along with different choices of $X_0$, are discussed in the Appendix, as illustrated in Figs.~\ref{fig10:appendix}(a)-\ref{fig10:appendix}(c).

\subsection{Suppression of Wigner negativity}\label{subsec:neg_decay}

Thermal decoherence from the heat reservoir causes the nonclassical interference fringes to disappear from the phase space of our compass states, as evidenced by the reduction of negative amplitudes of the associated Wigner distributions. Over time, this process transforms the Wigner function into a positively peaked distribution, indicating the transition of the quantum state into a classical one. This transition is clearly reflected in the temporal evolution of both the Wigner distribution and the tomogram functions discussed in the previous section. 

\begin{figure}[htp!]
\centering
\includegraphics[width=0.45\textwidth]{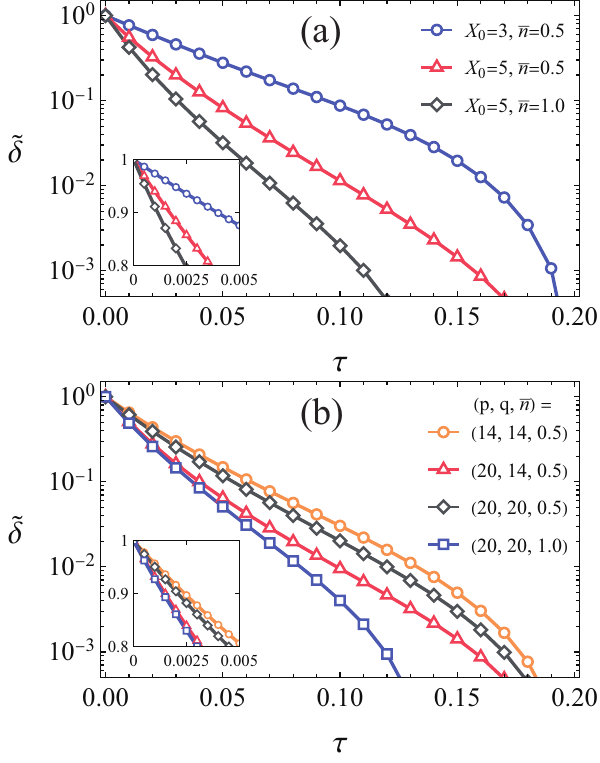}
\caption{The temporal evolution of the Wigner negativity of the compass states under the effects of the thermal reservoir: (a)~$p = q = 0$ and (b) $X_0 = 1.5$.}
\label{fig:fig5}
\end{figure}

The decay of nonclassical features in the states can be further investigated by analyzing the temporal evolution of Wigner negativity. In our study, the temporal evolution of Wigner negativity is numerically evaluated by employing
\begin{align}\label{eq:neg_vol2}
     \tilde{\delta}(\tau)= \frac{\int |W_{\ket{\ltimes}}(\beta,\tau)|\mathrm{d}^2\beta-1}{\int |W_{\ket{\ltimes}}(\beta,0)|\mathrm{d}^2\beta-1},
\end{align}
which quantifies the ratio of the Wigner negativity of the state after interaction with the reservoir to that of the initial state. This normalization ensures $\tilde{\delta}(0) = 1$ for all states, thereby enabling consistent comparison across different quantum states, and whose graphical illustration is provided in Fig.~\ref{fig:fig5}, evidencing the decay of the Wigner negativity over time with variations of involved parameters.

Figure~\ref{fig:fig5}(a) depicts the temporal evolution of the Wigner negativity of the standard compass state ($p = q = 0$) by varying $X_0$ and $\overline{n}$. The blue and red curves show that increasing $X_0$ causes a greater loss of negativity in the Wigner function. Moreover, the increase in the average thermal photon $\overline{n}$, as shown by the red and black curve in Fig.~\ref{fig:fig5}(a), demonstrates that the negativity in the Wigner function decreases quickly with higher values of $\overline{n}$.

Figure \ref{fig:fig5}(b) illustrates the impact of decoherence on the optimal compass state by changing the number of photons added ($p$) and subtracted ($q$), and by utilizing varied quantities of thermal photons $\overline{n}$, while keeping the amplitude $X_0$ constant. As depicted by the transition from the orange to the red curve via increasing $p$, and from the red to the black curve by varying $q$, along with the change from the black to the blue curve for different values of $\overline{n}$, the following trends emerge: Higher photon addition $p$ accelerates the loss of Wigner negativity, while photon subtraction $q$ helps preserve it in our adopted cases. Same as in the previous case of the compass state, an increase in the average thermal photon $\overline{n}$ leads to a greater loss of Wigner negativity corresponding to the optimized compass state. We further analyze the temporal evolution of the Wigner negativity for states with smaller values of $p$ and $q$, using multiple values of $X_0$, in the Appendix, as shown in Figs.~\ref{fig10:appendix}(d)-\ref{fig10:appendix}(f). The results show consistent behavior of Wigner negativity across our multiple parameter settings.

\subsection{Loss of quantum purity}\label{subsec:entropy}

The loss of quantum features from the phase space associated with the compass states, resulting from interaction with the heat reservoir and their eventual transition into a thermal state, can be confirmed by the degradation of their purity, as discussed in this section. The purity sustained by a quantum state $\hat{\rho}(\tau)$ over time $\tau$ can be assessed using $\text{tr}\hat{\rho}^2(\tau)$. This enables the classification of pure states with $\text{tr}\hat{\rho}^2(\tau) =1$ and mixed states with $\frac{1}{N}\le \text{tr}\hat{\rho}^2(\tau) < 1 $, where $N$ is the dimension of the density operator.

\begin{figure}
\centering
\includegraphics[width=0.45\textwidth]{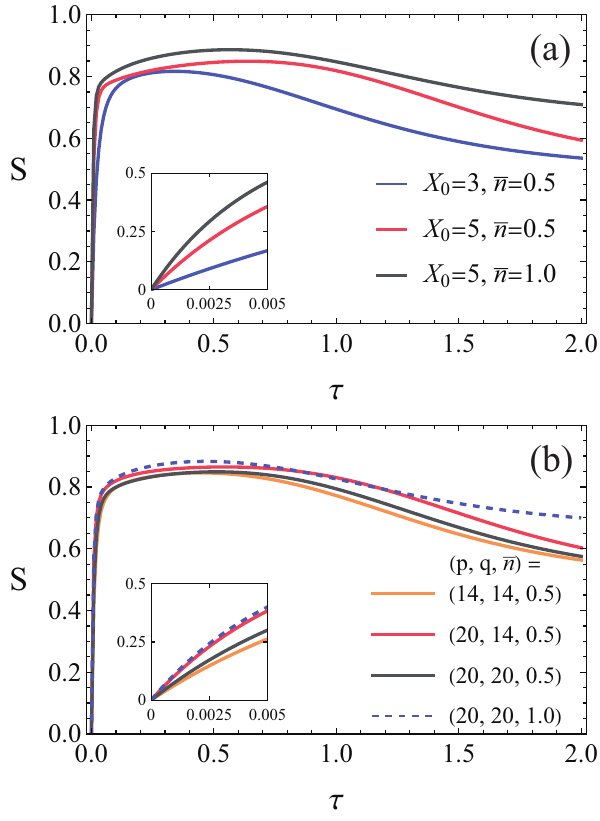}
\caption{Temporal evolution of the linear entropy of the compass states under the influence of the thermal reservoir, with insets showing shorter time intervals: (a)~$p = q = 0$ and (b)~$X_0 = 1.5$.}
\label{fig:fig6}
\end{figure}

Alternatively, the linear entropy, which quantifies the mixedness (or purity), is defined as $S(\tau) = 1 - \text{tr}[\hat{\rho}^2(\tau)]$. In terms of the Wigner function $W$, it is given by~\cite{PhysRevA.83.062121}
\begin{align}\label{eq:entropy}
    S(\tau)=1-\pi \int \mathrm{d}^2\beta W^2(\beta,\tau).
\end{align}
We numerically evaluate this equation and provide a graphical representation of each case, as shown in Fig.~\ref{fig:fig6}. Across each depicted case, we also carried out a detailed numerical investigation under the variation of the different parameters, which is enclosed in Table \ref{t1}. 

When $\tau=0$, the quantum states are pure, indicating that the linear entropy is given by $S(0) = 0$. As the coupling of the states with the heat reservoir is significant, the quantum states gradually lose their quantum nature and completely transform into a thermal state. The resulting thermal state $\hat{\rho}_\mathrm{th}$ can be denoted as
\begin{equation}
\hat{\rho}_\mathrm{th} = \sum_{m=0}^{\infty} \frac{\overline{n}^m}{(1 + \overline{n})^{m+1}} \ket{m}\bra{m},
\end{equation}
where $\ket{m}$ represents the eigenstates. At the long time limit ($\tau \to \infty$), the system reaches the thermal state $\hat{\rho}_\mathrm{th}$, and the linear entropy approaches a constant value $S(\tau \to \infty) = \frac{2\overline{n}}{2\overline{n}+1}$, which is characteristic of the thermal equilibrium state.

The rate of change of entropy ($\dot{S}$) can be expressed as,
\begin{equation}\label{eq:dsdt}
\dot{S}(\tau) = \frac{\mathrm{d}S}{\mathrm{d}\tau},
\end{equation}
and after a long interaction $(\tau \to \infty)$ with the heat reservoir, the system reaches thermal equilibrium, a steady state in which the entropy reaches a constant level for which $\dot{S}(\tau \to \infty) = 0$. However, in contrast to the long-term interaction with the heat reservoir, the change in the entropy of the compass states in the vicinity of the $\tau = 0$ region, denoted as $S_0 = \dot{S}(0)$, clearly reflects the impact of decoherence on the states. In other words, $S_0$ represents the instantaneous rate of change in the entropy, which we evaluate under different parameter choices by analyzing the relative variations in our depicted instances~\cite{vartia1976relative}.

The relative change in the entropy between two comparable cases is obtained by using
\begin{equation}
\Delta=\Bigg(\frac{S^B_0-S^A_0}{|S^A_0|}\Bigg)\times 100,
\end{equation}
which quantifies the purity differences between two comparable situations, $A$ and $B$, as an increase (for $\Delta >0$) or a decrease (for $\Delta <0$). We provide a graphical representation of multiple situations shown in Fig.~\ref{fig:fig6}, followed by a numerical investigation provided in Table \ref{t1}.

Starting with $p = q = 0$, depicted in Fig.~\ref{fig:fig6}(a), which represents the compass state when $X_0$ is greater. First, let us check the impact of the amplitude $X_0$ on the rate of change of entropy, which is observed along the blue and red curves of Fig.~\ref{fig:fig6}(a), implying that as $X_0$ is increased by keeping other parameters constant, the red curve with the higher amplitude $X_0$ shows a steeper slope near $\tau = 0$ compared to the blue curve, which is quite apparent at the shorter range of time as shown in the inset of Fig.~\ref{fig:fig6}(a). This demonstrates that as the parameter $X_0$ is increased, the entropy attains a nonzero value more quickly, as evidenced by the rate of change of entropy for the blue curve, $S_0 = 38.945$, and for the red curve, $S_0 = 101.999$. The numerical results of these two cases are illustrated in Table \ref{t1} along the rows $i=1$ and $i=2$, for which the relative rise in the entropy is measured as $\Delta=161.90 \%$, implying that in this situation, an increase in the $X_0$ from $X_0=3$ to $X_0=5$ enhances the loss in purity up to around $161.90\%$.

An increase in the average thermal photon number of the reservoir leads to a larger change in entropy, as can be observed from the red and black curves in Fig.~\ref{fig:fig6}(a). This behavior becomes even clearer when the two curves are compared over a short interval of time, as shown in the inset of Fig.~\ref{fig:fig6}(a). The entropy change for the black curve is found to be $S_0 = 153.999$ ($i = 3$ in Table~\ref{t1}), which is greater than the value for the red curve, $S_0 = 101.999$ ($i = 2$ in Table~\ref{t1}). These values provide numerical evidence of the stronger increase in entropy as the average photon number increases.

Consider the cases under the change of $p$, $q$, and $\overline{n}$, while keeping $X_0$ at a constant lower value, as shown in Fig.~\ref{fig:fig6}(b) and their numerical investigation is presented in Table \ref{t1} from $i=4$ to $i=7$. First, consider the two comparable cases presented along the rows $i=4$ and $i=5$ in Table \ref{t1}, where it is shown that increasing the quantity of added photons leads the optimized compass state to lose its purity quickly, which is evident from the orange and red curves with lower and higher photon ($p$) values, respectively. Specifically, for the orange curve, $S_0= 66.760$, while for the red curve $S_0 = 112.563$, providing clear numerical evidence for this trend, which further can also be confirmed through the curves provided in the inset of Fig.~\ref{fig:fig6}(b) for a relatively short value of $\tau$ and the relative increase in the linear entropy measured through comparing $S_0$ for these two cases is $\Delta=68.61\%$.

In contrast to the photon addition, we observe that a higher amount of photon subtraction from the states slows the decay rate of the purity, which is evident in the red and black curves shown in the inset of Fig.~\ref{fig:fig6}(b). For the black curve, we evaluate $S_0 = 79.784$, which is lower than that of the red curve, indicating that increasing the amount of subtracted photons ($q$) from the optimized compass state decreases the entropy change. These two cases are presented along $i=5$ and $i=6$ in Table \ref{t1}, implying that increasing photon subtraction ($q=14$ to $q=20$) from the optimized compass state results in a relative drop in entropy change of about $\Delta=29.12\%$. 

Finally, the increase in the average thermal photon number (compare $i=6$ and $i=7$ in Table \ref{t1}) also leads to a rise in the linear entropy, as both graphically and numerically verified in these instances. This is evident in the comparison between the black and blue curves with lower and higher $\overline{n}$ values, respectively, which is then clearly visible in the inset of Fig.~\ref{fig:fig6}(b). For the blue curve we have $S_0= 120.676$, indicating a greater slope compared to the black curve, which corresponds to a lower value of $\overline{n}$ and exhibits a slower rate of entropy change, and for this case the relative increase in the entropy while increasing $\overline{n}$ is $\Delta=51.25\%$. The linear entropy of our states is further evaluated for different parameter values, showing that the observed trends remain preserved across different parametric settings, as shown in Figs.~\ref{fig10:appendix}(g)-\ref{fig10:appendix}(i).
\begin{table}[ht]
\centering
\caption{The values are obtained by numerical investigation of the cases presented in Fig.~\ref{fig:fig6}.}
\begin{tabular}[t]{lcccccc}
\toprule
\toprule
$i $ &\hspace{0.5cm} $X_0$ &\hspace{0.5cm} $p$ &\hspace{0.5cm} $q$&\hspace{0.5cm} $\overline{n}$ &\hspace{0.5cm}$S_0$&\hspace{0.5cm}$\Delta\%$ \\
\midrule
1&\hspace{0.5cm}3&\hspace{0.5cm}0&\hspace{0.5cm}0&\hspace{0.5cm}0.5&\hspace{0.5cm}38.945&\hspace{0.5cm}\XSolid \\
\midrule
2&\hspace{0.5cm}5&\hspace{0.5cm}0&\hspace{0.5cm}0&\hspace{0.5cm}0.5&\hspace{0.5cm}101.999&\hspace{0.5cm}161.90\\
\midrule
3&\hspace{0.5cm}5&\hspace{0.5cm}0&\hspace{0.5cm}0&\hspace{0.5cm}1&\hspace{0.5cm}153.999&\hspace{0.5cm}50.98\\
\midrule
4&\hspace{0.5cm}1.5&\hspace{0.5cm}14&\hspace{0.5cm}14&\hspace{0.5cm}0.5&\hspace{0.5cm}66.760&\hspace{0.5cm}\XSolid\\
\midrule
5&\hspace{0.5cm}1.5&\hspace{0.5cm}20&\hspace{0.5cm}14&\hspace{0.5cm}0.5&\hspace{0.5cm}112.563&\hspace{0.5cm}68.61\\
\midrule
6&\hspace{0.5cm}1.5&\hspace{0.5cm}20&\hspace{0.5cm}20&\hspace{0.5cm}0.5&\hspace{0.5cm}79.784&\hspace{0.5cm}-29.12\\
\midrule
7&\hspace{0.5cm}1.5&\hspace{0.5cm}20&\hspace{0.5cm}20&\hspace{0.5cm}1&\hspace{0.5cm}120.676&\hspace{0.5cm}51.25\\
\bottomrule
\end{tabular}
\label{t1}
\end{table}

\section{GENERALIZATION TO ARBITRARY STATES} \label{sec:phase_robustness1}

In the previous sections, two compass states constructed using different methods were shown to exhibit distinct phase-space characteristics, and their robustness against a thermal reservoir differed accordingly. These observations, supported by multiple theoretical approaches, underscore a strong phase space dependence in the robustness of quantum states. In this section, we investigate this dependence in a more general framework applicable to any arbitrary pure quantum state.

Let us begin by reformulating Eq.~(\ref{eq:models_decoh}) as a partial differential equation in terms of the Wigner function ($W$)
\begin{align} \label{eq:rescale_wig_master}
			\frac{\partial}{\partial t} W\left(x,y;t\right) = \omega \left(2 W + \bm{r} \cdot  \nabla W \right)+\frac{\omega}{2}\left(2\overline{n} +1\right) \nabla \cdot \nabla W,
\end{align}
where $\bm{r} = (x, y)^{\top}$ represents the radius vector in phase space and $\nabla = \left( \frac{\partial}{\partial x}, \frac{\partial}{\partial y} \right)$ denotes the phase space divergence operator, defined with respect to the position coordinate $x$ and the momentum coordinate $y$. The corresponding equation takes the form of the famous Fokker–Planck equation, with the two terms on the right-hand side of Eq.~(\ref{eq:rescale_wig_master}) illustrating distinct and intuitive physical interpretations.

The first term of Eq.~(\ref{eq:rescale_wig_master}) governs the contraction of structures in phase space, accompanied by an increase in the amplitude of the Wigner function; however, this process alone neither diminishes nor enhances the negativity of the Wigner function $(\delta)$ defined in Eq.~(\ref{eq:neg_vol2}). The second term of Eq.~(\ref{eq:rescale_wig_master}) corresponds to a diffusion process analogous to the well-known heat equation, through which the Wigner function redistributes from regions of higher amplitude to those of lower amplitude. As a result, the negativity of the Wigner function is gradually suppressed, leading to the eventual disappearance of nonclassical features. This also explains the behavior of $d(\tau)$ shown in the  Fig.~\ref{fig:fig4}, the nonclassical features localized near the origin of phase space are first suppressed via spatial averaging induced by an analogous to thermal diffusion, driving $d(\tau)$ toward zero. Subsequently, the contraction becomes dominant, causing $d(\tau)$ to increase and asymptotically approach the value characteristic of a thermal state.

\subsection{Features sizes and decoherence}

We first investigate decoherence from the heat reservoir on a quantum phase-space feature, linking its volume or area to the decay rate. To do so, we quantitatively analyze the effects on a phase-space structure by examining the instantaneous rate of change along a selective volume in the Wigner phase space. We consider a time-dependent domain $A(\tau)$ with boundary $\partial A(\tau)$. The volume of the Wigner function over this domain  is defined as $v(\tau)$, and thus the rate of change of $v(\tau)$ is defined as
\begin{equation}
\dot{v} \left( \tau \right) = \frac{\mathrm{d}  v \left( \tau \right)  }{\mathrm{d} \tau} =   \iint_{A \left( \tau \right) } \frac{\partial W \left(x,y;\tau \right)  }{\partial \tau} \mathrm{d} x \mathrm{d} y,
\end{equation}
when the Wigner function within the patch is positive, $v(\tau) > 0$; when it is negative, $v(\tau) < 0$. By applying the Reynolds transport theorem~\cite{walter2020classical}, we obtain 
\begin{equation}\label{eq:volume_2}
\dot{v} \left( \tau \right) = \left(\overline{n}+\frac{1}{2} \right) \oint_{\partial A\left( \tau \right)}  \nabla W \left(x,y;\tau\right) \cdot \bm{n} \mathrm{d}l,
\end{equation}
where $\bm{n}$ is the outward-pointing normal vector of the boundary $\partial A(\tau)$. Since the signs of $v(\tau)$ and any $\nabla W(x,y;\tau) \cdot \bm{n}$ on the boundary $\partial A(\tau)$ must be opposite, the tiny-scale structures in phase space are inevitably prone to disruption.

Apart from volume loss, we now examine the variation in the impact of decoherence across phase-space patches at the different scales (or areas), which can be characterized by tracking the motion of the zero points of the associated Wigner functions. Consider a simply connected region labeled as $A(\tau)$ with boundary $\partial A(\tau)$. The time evolution of the area $a_{\pm}(\tau)$ of this region is given as follows,
\begin{equation}\label{eq:neg_area}
	\dot{a}_{\pm} \left( \tau \right) =\frac{\mathrm{d}a_{\pm}\left( \tau \right)}{\mathrm{d}{\tau}} = \frac{\mathrm{d}}{\mathrm{d}{\tau}} \iint_{A\left( \tau \right)} \mathrm{d}x  \mathrm{d}y = \int_{\partial A\left( \tau \right)} \bm{v}_{b} \cdot \bm{n} \mathrm{d}l,
\end{equation}
where the subscript of $a_{\pm}$ indicates whether the patch is positively ($+$) or negatively peaked ($-$), $\bm{v}_{b} $ is the Eulerian velocity of the boundary, which should not be confused with the velocity field associated with the Wigner function~\cite{oliva_anharmonic_2018}.

The boundary conditions $W\left(x(\tau),y(\tau);\tau\right)=0$ in the Lagrangian description lead to
\begin{equation}
 \frac{\mathrm{d} W }{\mathrm{d} \tau} = 	\frac{\partial} {\partial \tau }W + \left(\frac{\partial}{\partial{\tau}} \bm{r}\right) \cdot \nabla W = 0,
\end{equation}
and with the Lagrangian and Eulerian specifications of the flow field, which are related as $\bm{v}_{b} = \frac{\partial \bm{r}}{\partial{\tau}} $. Equation (\ref{eq:neg_area}) is rewritten as follows:
\begin{align}\label{eq:neg_area2new}
\dot{a}_{\pm} \left( \tau \right) =&\nonumber - 2 a_{\pm} \left( \tau \right)
  \pm \left(\overline{n}+\frac{1}{2} \right)\oint_{\partial A(\tau)} \frac{\nabla \cdot \nabla W}{\rvert \nabla W \lvert} \mathrm{d} l\\
  =& \nonumber - 2 a_{\pm} \left( \tau \right) -2\pi\left(\overline{n}+\frac{1}{2} \right) \\&-\left(\overline{n}+\frac{1}{2} \right) \oint_{\partial A(\tau)}  \nabla\left( \ln \lvert \nabla W \rvert \right) \cdot \bm{n} \mathrm{d} l.
\end{align}
Here, $-2a_{\pm}\left( \tau \right)$ arises from a term associated with the contraction of the phase space pattern in the Eq.~{(\ref{eq:rescale_wig_master})}, while $-2\pi\left(\overline{n}+\frac{1}{2} \right)$ represents a geometric effect~\cite{goldman_curvature_2005} independent of the specific shape of the boundary $\partial A(\tau)$.

\begin{table}[ht]
	\centering
	\caption{The values are obtained through the numerical integration of Eqs.~(\ref{eq:volume_2}) and (\ref{eq:neg_area2new}) for the specific instances of the compass state and its optimized version, as considered in our work.}
	\begin{tabular}[t]{lcccccccc}
		\toprule
		\toprule
		$i $ &\hspace{0.2cm} $X_0$ &\hspace{0.2cm} $p$ &\hspace{0.2cm} $q$ &\hspace{0.2cm}$v(0)$&\hspace{0.2cm}$\dot{v}(0)$ &\hspace{0.2cm}$\frac{\dot{v}(0)}{v(0)}$  &\hspace{0.2cm} $a_{+}(0)$&\hspace{0.2cm}$\dot{a}_{+}(0)$ \\
		\midrule
		1&\hspace{0.2cm}3&\hspace{0.2cm}0&\hspace{0.2cm}0&\hspace{0.2cm}0.129&\hspace{0.2cm}-4.52 &\hspace{0.2cm}-35.0&\hspace{0.2cm}0.503&\hspace{0.2cm}1.68\\
		\midrule
		2&\hspace{0.2cm}5&\hspace{0.2cm}0&\hspace{0.2cm}0&\hspace{0.2cm}0.0500&\hspace{0.2cm}-4.80&\hspace{0.2cm}-96.1&\hspace{0.2cm}0.197&\hspace{0.2cm}1.18\\
		\midrule
		3&\hspace{0.2cm}1.5&\hspace{0.2cm}14&\hspace{0.2cm}14&\hspace{0.2cm}0.0739&\hspace{0.2cm}-4.88&\hspace{0.2cm}-65.9&\hspace{0.2cm}0.274&\hspace{0.2cm}0.0918\\
		\midrule
		
        4&\hspace{0.2cm}1.5&\hspace{0.2cm}20&\hspace{0.2cm}20&\hspace{0.2cm}0.0624&\hspace{0.2cm}-4.87&\hspace{0.2cm}-78.1&\hspace{0.2cm}0.234&\hspace{0.2cm}0.300\\
        \midrule
        
        5&\hspace{0.2cm}1.5&\hspace{0.2cm}20&\hspace{0.2cm}14&\hspace{0.2cm}0.0441&\hspace{0.2cm}-4.96&\hspace{0.2cm}-112&\hspace{0.2cm}0.163&\hspace{0.2cm}-0.0545\\

		\bottomrule
	\end{tabular}
	\label{t2}
\end{table}

In a special case, if the patch $A(\tau)$ under examination is a circular-shaped region with a radius $r$ and the Wigner function has rotational symmetry along its boundary. The coordinate system centered at the origin, defined as the center of the circle, has a simple form
\begin{equation}
	\dot{a}_{\pm} \left( \tau \right) = - 2 a_{\pm} \left( \tau \right) -2\pi\left(\overline{n}+\frac{1}{2} \right) \left(1+ r \frac{W^{\prime\prime}(r)}{W^\prime(r)}\right).
\end{equation}
In Table \ref{t2}, we analyze the decay of the central phase-space feature of the associated compass state and its optimized variant. We computed the area $a_{+}(0)$ of the central patch (among all the enumerated examples, the considered patches are positively peaked) and the volume of the Wigner function $v(0)$ over the patch, as well as the instantaneous rates of change of both the area $\dot{a}_{+}(0)$ and the volume $\dot{v}(0)$, evaluated at an average thermal photon number kept constant to $\overline{n}=0.5$.

From Table \ref{t2}, the relative instantaneous reduction in the area denoted as $\dot{a}_{+}(0)$ and the relative reduction in the volume measured through $\frac{\dot{v}(0)}{v(0)}$ highlight the decreases across different initial patches of areas $a_{+}(0)$ and volumes $v(0)$ in the presence of thermal decoherence. These results show that patches at lower scales tend to lose their existence more quickly. This behavior is further corroborated by our analysis in the previous section, where it is justified with different theoretical frameworks. It is observed that a reduction in the size of the smallest central phase space features leads to an enhanced susceptibility to decoherence, suggesting that quantum states at these finer scales are more vulnerable to decoherence.

We now employ a series of approximations to characterize the dependence of the decoherence rate on the properties of  the phase space patch $A(\tau)$. As in the preceding analysis, we model the patch as a circular region of radius $r$ in phase space. Under this assumption, Eq.~(\ref{eq:volume_2}) simplifies significantly as follows:
\begin{equation}\label{eq:dvdt}
	\dot{v}(\tau) = \left(2 \overline{n}+ 1 \right) \pi r W^\prime(r).
\end{equation}
On this circular patch, a generalized cone is defined with a circular directrix and curved generators. Its volume is given by $v(\tau) = \chi \pi r^2 h$, where $\chi$ is the geometric shape factor and $h$ denotes the height of the generalized cone (for a standard cone, $\chi = 1/3$). Applying the finite-difference method to approximate the radial derivative of the Wigner function at the boundary $\partial A(\tau)$, yielding $W'(r) \simeq \frac{W(r) - W(0)}{r} = -\frac{h}{r}$. Combining this with the volume expression in Eq.~(\ref{eq:dvdt}) we obtain the estimation formula
\begin{equation}
\frac{\dot{v}(\tau)}{v(\tau)}  \simeq  -\frac{2 \overline{n}+ 1}{\chi r^2} =  -\frac{(2 \overline{n}+ 1)\pi}{\chi a_{\pm}(\tau)},
\end{equation}
where $a_{\pm}(\tau) = \pi r^2$ is the area of the circular patch. Consequently, the magnitude of the relative volume decay rate $\left| \frac{\dot{v}(\tau)}{v(\tau)} \right|$ is approximately inversely proportional to $a_{\pm}(\tau)$: smaller patches decay faster. This result is directly related to the decay of the central structure as characterized by $d(\tau)$.

\subsection{Wigner negativity and entropy trend}

We next reformulate Eq.~(\ref{eq:neg_vol}) into a strictly equivalent expression, $\delta =(-1) \sum_{-} v$, based on its intrinsic definition, where $\sum_{-}$ denotes the summation over all negative patches in phase space. Consider the time derivative of the relative Wigner negativity $\tilde{\delta}(\tau)$ of the state in Eq.~(\ref{eq:neg_vol2}) evaluated at $\tau = 0$:
\begin{equation}\label{eq:inequal}
	 \frac{\mathrm{d}  }{\mathrm{d} \tau}  \tilde{\delta}(0) = \frac{\sum_{-}\dot{v}(0)}{\sum_{-}v(0)} \ge -\max\left| \frac{\dot{v}(0)}{v(0)} \right|\ge \sum_{-}\frac{\dot{v}(0)}{v(0)},
\end{equation}
where we have applied the Cauchy inequality. Here $\max\left| \frac{\dot{v}(0)}{v(0)} \right|$ is the maximum value of the relative volume decay rate over all negative patches. Eq.~(\ref{eq:inequal}) indicates that the smallest structure in phase space limits the overall decoherence rate. In conclusion, our analysis establishes a quantitative connection between the non‑classicality of the state, as measured by Wigner negativity, and the characteristic scales in phase space.

Finally, we clarify the relation between the entropy production rate in Eq.~(\ref{eq:dsdt}) and the phase space structure. To this end, substitute Eq.~(\ref{eq:rescale_wig_master}) into Eq.~(\ref{eq:entropy}).  Noting that $\mathrm{d}x\,\mathrm{d}y = 2\mathrm{d}^2\beta$, the entropy production rate becomes
\begin{equation}\label{eq:dsdt2}
\dot{S}(\tau) = 2\left[S(\tau)-1\right] + \frac{\pi}{2}(2 \overline{n}+ 1) \iint\lvert \nabla W\rvert^2 \mathrm{d}x \mathrm{d}y.
\end{equation}
This indicates that the change in entropy can be attributed to the Dirichlet energy of the Wigner function $W(x,y)$, which appears in the second term in Eq.~(\ref{eq:dsdt2}). The Dirichlet energy measures the total squared rate of change of a function over a given domain~\cite{Rudin1987}, which for a real-valued function $g(x,y)$ defined on a domain $\Omega \subseteq \mathbb{R}^2$, it is mathematically defined as $E_\Omega[g] = \iint_\Omega |\nabla g(x,y)|^2 \, \text{d}x \, \text{d}y$, with $\nabla g= \left(\frac{\partial g}{\partial x}, \frac{\partial g}{\partial y} \right)$ being the gradient of $g$. This implies that finer scales, more rapidly varying structures in phase space, directly amplify the entropy production rate $\dot{S}(\tau) $.

We now further elucidate the relationship between the entropy production rate and the patches $A(\tau)$. On each patch, the boundary $\partial A(\tau)$ satisfies $W=0$, thus we invoke the Rayleigh–Ritz variational principle for the Laplacian $\Delta = -\left(\frac{\partial^2}{\partial x^2} + \frac{\partial^2}{\partial y^2}\right)$ on the patch $A(\tau)$:
\begin{equation}\label{eq:ineq1}
\lambda_{1}(A)  \le \frac{\iint_{A(\tau)} W \Delta W \mathrm{d}x \mathrm{d}y }{\iint_{A(\tau)} W^2  \mathrm{d}x \mathrm{d}y},
\end{equation}
and by integration by parts, this inequality can also be written as
\begin{equation}
\iint_{A(\tau)} |\nabla W|^2  \mathrm{d}x \mathrm{d}y \ge \lambda_{1}(A) \iint_{A(\tau)} W^2 \mathrm{d}x \mathrm{d}y,
\end{equation}
where $\lambda_{1}(A)$ is the lowest eigenvalue of $\Delta$ on the domain $A(\tau)$, also known as the first Dirichlet eigenvalue of the patch.
The Faber–Krahn inequality states that, among all patches of a given area, the disk minimizes the first Dirichlet eigenvalue. Hence, for a patch of area $a_{\pm}(\tau)$,
\begin{equation}\label{eq:ineq2}
\lambda_{1} (A) \ge \frac{\pi j_{0,1}^2}{ a_{\pm}(\tau)},
\end{equation}
where $j_{0,1} \simeq 2.405$ is the first positive root of Bessel function of the first kind of order zero $\mathrm{J}_{0}(x)$. Cauchy–Schwarz inequality  allow us to directly relate the right side of the Eq.~(\ref{eq:ineq1}) to the properties of the patch
\begin{equation}\label{eq:ineq3}
\iint_{A(\tau)} W^2 \mathrm{d}x \mathrm{d}y \ge \frac{v^2(\tau)}{a_{\pm}(\tau)}.
\end{equation}
Combining these results gives a lower bound on the entropy production rate in terms of the phase space patches,
\begin{equation}\label{eq:dsdt3}
\dot{S}(\tau) \ge 2\left[S(\tau)-1\right] + \frac{\pi^2 j_{0,1}^2 }{2}(2 \overline{n}+ 1) \sum_{\pm} \left[ \frac{v(\tau)}{a_{\pm}(\tau)}\right]^2,
\end{equation}
where $\sum_{\pm}$ denotes the summation over all patches in phase space. This results align with the  expectations derived from Eq.~(\ref{eq:dsdt2}): the lower bound of entropy production rate increases with both the number of patches and the peak value of each patch.

In summary, we have established a quantitative link between several key signatures of decoherence [e.g., the central structure characterized by $d(\tau)$, the instantaneous decay rate of relative Wigner negativity $\frac{\mathrm{d}  }{\mathrm{d} \tau} \tilde{\delta}(0)$ , and the entropy production rate $\dot{S}(\tau)$] and the phase space structure, independent of the specific functional form of the Wigner function. These results are specifically confined to a single-mode optical field $W(x,y)$, which may also be extendable to multi-mode optical fields.

\section{SUMMARY}\label{sec:conc}

We considered the compass state and its optimized variant obtained by applying photon addition and then subtraction operations, and both of these two states provide a variety of phase-space features with comparatively larger extensions to the tiny features at the sub-Planck scale, as well as a fair comparison of the varying effects of decoherence across different phase-space scales. We notice that photon addition and then subtraction on the compass state can yield an isotropic form of the sub-Planck structures in the same fashion as applying photon subtraction first and then photon addition on the compass state, as discussed in the previous work~\cite{akhtar2024sub-shot}. Note that the current variant of the compass state was overlooked to exhibit isotropy in its sub-Planck phase-space structures and sensitivity to perturbations, a positive outcome for quantum sensing. 

We mainly focused on the impact of decoherence arising from a heat bath on the representative compass states, supported by the number of theoretical techniques resulting in unified results. We established a connection between the effects of decoherence and the spatial sizes or extents of the phase-space structures associated with the outlined compass states. Our findings are generally applicable to any arbitrary pure quantum state interacting with the considered heat reservoir. Our findings indicate that quantum states with finer phase-space structures are more prone to decoherence, which is justified through our representative compass states. In the cases of compass states we examined, adding photons, or increasing the amplitudes of the corresponding superposed component coherent states, results in progressively delicate phase-space features, making the states more fragile. Photon subtraction produces the opposite effect. Small-scale structures degrade faster than those occupying larger regions of phase space, highlighting the limited robustness of crucial sub-Planck structures when decoherence is present.

\section*{ACKNOWLEDGEMENT}
This work was supported by the National Natural Science
Foundation of China (Grants No.~12475009, No.~12075001,
and No.~61601002), the Anhui Provincial Key Research and Development Plan (Grant No.~2022b13020004), the Anhui Province
Science and Technology Innovation Project (Grant No.
202423r06050004), the Anhui Provincial Department of Industry and Information Technology (Grant No.~JB24044), the Anhui Province Natural Science Foundation (Grant No. 2508085ZD001), and the Anhui Provincial University Scientific Research Major Project (Grant No.~2024AH040008). J.-X Peng acknowledges the support from the National Natural Science Foundation of China (Grant No.~12504566), the Natural Science Foundation of Jiangsu Province  (Grant No.~BK20250947), the Natural Science Foundation of the Jiangsu Higher Education Institutions (Grant No.~25KJB140013), and  the Natural Science Foundation of Nantong City (Grant No.~JC2024045).
\vspace{0cm}
\appendix

\section{Intermediate photon-varied states}
\begin{figure}[htp!]
\centering
\includegraphics[width=0.45\textwidth]{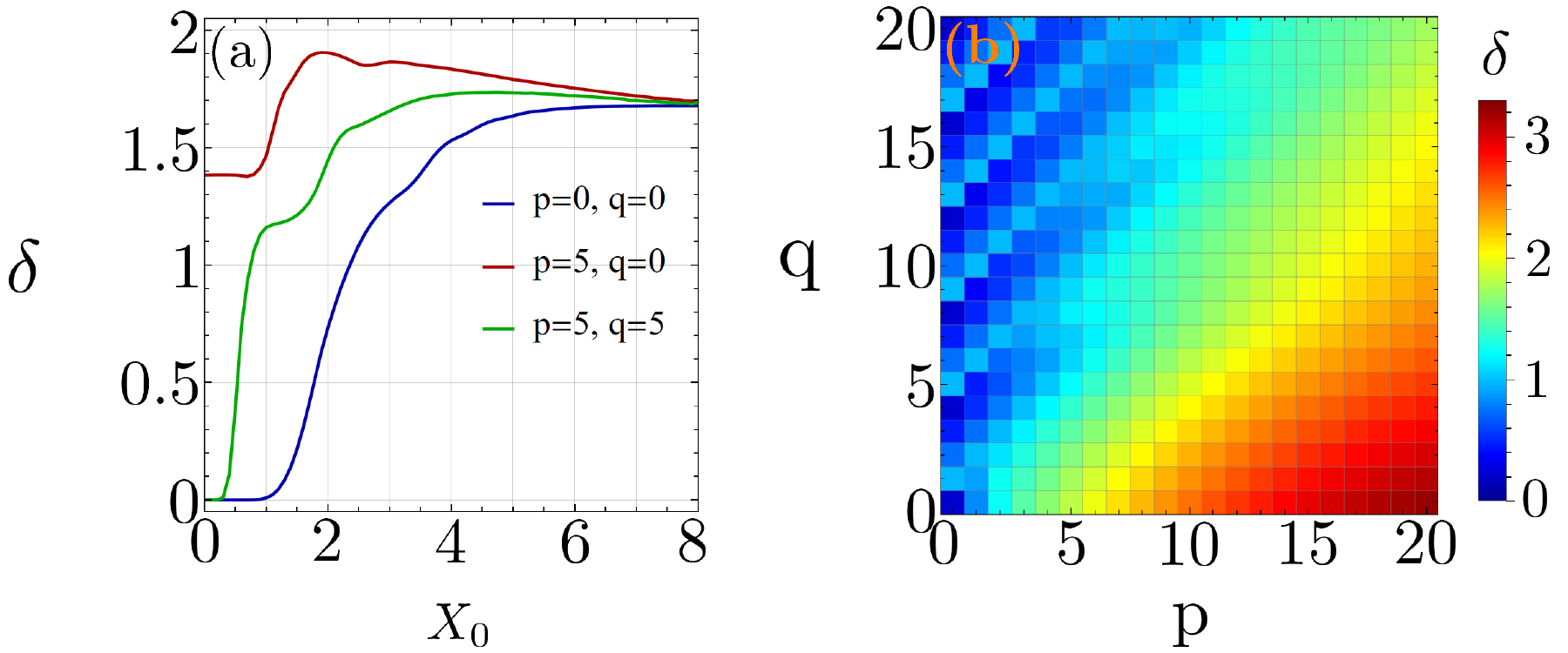}
\caption{Wigner negativity $W_{\ket{\ltimes}}(\beta)$: (a) $p$ and $q$ are fixed while $X_0$ is varied; (b) $X_0 = 1.5$ is fixed while $p$ and $q$ are varied.}
\label{fig1:appendix}
\end{figure}
Here, we extend our demonstrations to include scenarios where intermediate photon addition and subtraction operations are explored, demonstrating that our findings hold true regardless of the specific parameter choices for our quantum states.

We first examine the Wigner negativity of our states prior to interaction with the specified heat reservoir, as shown in Fig.~\ref{fig1:appendix}. For the case with a fixed $p=q=0$ [blue curve in Fig.~\ref{fig1:appendix}(a)] and for nonzero photon addition and subtraction values (red curve: $p=5, q=0$; green curve: $p=q=5$), the Wigner negativity initially increases with $X_0$ and then reaches a constant value for larger $X_0$ values. This behavior aligns with the trend observed for higher photon addition and subtraction values in Fig.~\ref{fig:fig2}, indicating that the pattern persists from intermediate to higher photon operations.

Let us now consider the cases where $X_0$ is fixed and $p$ and $q$ are varied simultaneously [Fig.~\ref{fig1:appendix}(b)]. This shows that for comparatively higher photon addition and subtraction values, increasing $p$ enhances the Wigner negativity, whereas increasing $q$ reduces it. For zero photon addition (along the $q$ axis), the state returns to the original form at multiples of 4, being an eigenstate of $\hat{a}^4$ as discussed in the main section; in other cases, it corresponds to a phase-shifted state.
\begin{figure}
\centering
\includegraphics[width=0.5\textwidth]{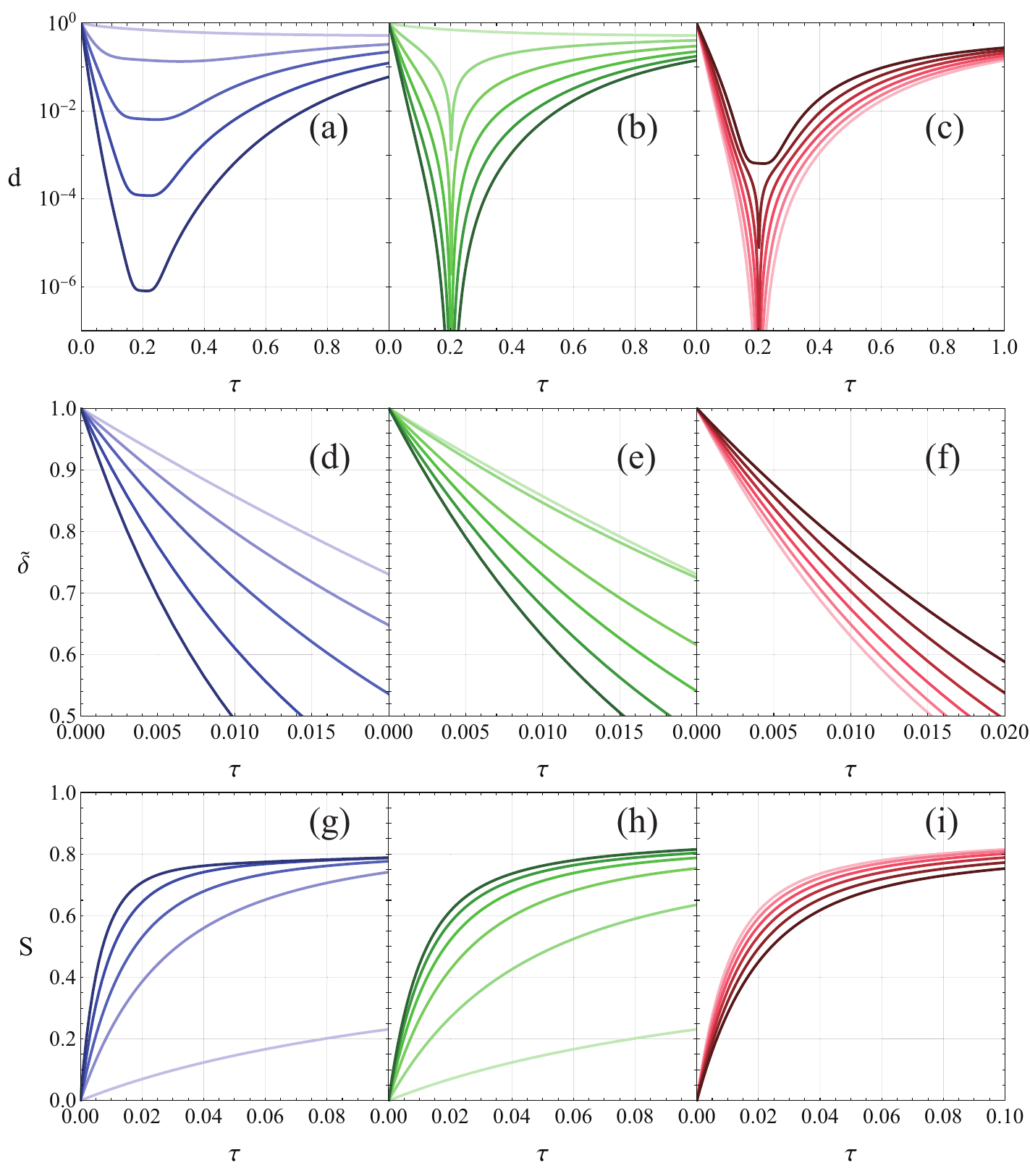}
\caption{Blue curves correspond to the case with $p = q = 0$, and $X_0$ is varied over $1.5, 2.5, 3.5, 4.5, 5.5$; the curve darkens monotonically as $X_0$ increases. Green curves show $X_0 = 1.5$ and $q = 0$, with $p$ varied over $0, 1, 2, 3, 4, 5$, darkening monotonically with increasing $p$. Red curves represent the case when $X_0 = 1.5$ and $p = 5$, with $q$ variation over $0, 1, 2, 3, 4, 5$ and darkening monotonically as $q$ increases. In all cases, $\bar{n} = 0.5$. (a)–(c) depict the decay of central peaks, (d)–(f) show the temporal evolution of Wigner negativity, and (g)–(i) display the corresponding evolution of linear entropy.}
\label{fig10:appendix}
\end{figure}
We now analyze cases in which decoherence is present, focusing on its impact on the central phase-space features, Wigner negativity, and purity under moderate parameter selections, as shown in Fig.~\ref{fig10:appendix}. When $p$ and $q$ are fixed and $X_0$ are varied over $1.5, 2.5, 3.5, 4.5, 5.5$, the blue curve darkens monotonically with increasing $X_0$. For fixed $X_0$ and $q$, but $p$ is varied over $0, 1, 2, 3, 4, 5$, with the green curve darkening monotonically with greater $p$. Similarly, for $X_0$ and $p$ fixed to nonzero values, and $q$ is varied over $0, 1, 2, 3, 4, 5$, and the red curve darkens monotonically with increasing $q$. This indicates that the decay of the central peak, the reduction of Wigner negativity, and the loss of quantum purity due to decoherence are enhanced as $p$ (photon addition) and $X_0$ (coherent state amplitude) increase, whereas $q$ (photon subtraction) can have the opposite effect under fixed nonzero $p$ and $X_0$ values.
\section*{DATA AVAILABILITY}
There are no publicly available research data or software supporting this manuscript. Requests for further information
or data should be sent to the authors.
\bibliography{ref}
\end{document}